\documentclass[aps,prd,twocolumn,amsmath,amssymb,nofootinbib]{revtex4}
\usepackage{graphicx,color,dcolumn,booktabs}

\usepackage{graphicx}
\usepackage{array,color}
\usepackage{mathrsfs}
\usepackage{longtable,lscape}
\usepackage{txfonts}
\usepackage{amssymb}
\usepackage{indentfirst}
\usepackage{color}
\usepackage{amssymb}
\usepackage{epsfig}
\usepackage{multirow}
\usepackage{subfigure}
\newcommand{\veL}{\mbox{\boldmath${\rm L}$}}
\newcommand{\veS}{\mbox{\boldmath${\rm S}$}}
\usepackage[colorlinks=true,
            urlcolor=blue,
            linkcolor=blue,
            citecolor=blue,
            bookmarks=true,
            bookmarksopen=true,
            bookmarksnumbered=true]{hyperref}

\begin{document}

\title{{\bf Charmonium states in a coupled-channel model }}

\author{Zi-Long Man$^1$}\email[Electronic address:]{manzilong@mail.sdu.edu.cn}
\author{Cheng-Rui Shu$^1$}
\author{Yan-Rui Liu$^1$}\email[Electronic address:]{yrliu@sdu.edu.cn}
\author{Hong Chen$^2$}\email[Electronic address:]{chenh@swu.edu.cn}

\affiliation{$^1$School of Physics, Shandong University, Jinan, Shandong 250100, China\\
	$^2$School of Physical Science and Technology, Southwest University, Chongqing 400715, China}

\begin{abstract}
We systematically investigate the mass spectrum and two-body open-charm strong decays of charmonium states in a coupled-channel model where the $^3P_0$ quark-antiquark pair creation mechanism is employed. The results of masses, mass shifts, proportions of the $c\bar{c}$ component, and open-charm decay widths are provided. The $S$-$D$ wave mixing angles and di-electric decay widths for vector mesons are also presented. Based on our results, we find that the $\psi(3770)$, $\psi(4040)$, $\psi(4160)$, $\psi(4360)$, and $\psi(4415)$ can be assigned as the $1^3D_1$-, $3^3S_1$-, $2^3D_1$-, $4^3S_1$-, and $3^3D_1$-dominated charmonium states, respectively. The $\psi_3(3842)$ is a good candidate of the $\psi_3(1D)$ charmonium state. The calculated mass and strong decay width of $\chi_{c1}(2P)$ with significant continuum contribution ($\sim$57\%) favor the charmonium interpretation for the mysterious $\chi_{c1}(3872)$. When considering the large uncertainty in the observed decay width, the possibility to assign the $\chi_{c0}(3860)$ as the $\chi_{c0}(2P)$ charmonium state cannot be ruled out. One may describe well the properties of $\chi_{c2}(3930)$ with the $\chi_{c2}(2P)$ charmonium. The predictions on properties of other $c\bar{c}$ states can be tested by future experiments.

\end{abstract}

\date{\today}


\maketitle

\section{Introduction}\label{sec1}

The exotic state $\chi_{c1}(3872)$ was first observed in the exclusive $B^\pm\rightarrow K^\pm\pi^{+}\pi^{-}J/\psi$ decays by the Belle Collaboration in 2003 \cite{Belle:2003nnu}. After that, more and more charmonium and charmonium-like states, such as $\chi_{c0}(3915)$, $\chi_{c2}(3930)$, $\psi(4230)$, $\psi(4360)$, $\psi(4500)$, $\psi(4660)$, and $\psi(4700)$, were announced by the BaBar, Belle, BESIII, D0, CDF, CMS, and LHCb Collaborations \cite{ParticleDataGroup:2022pth}. In 2017, the charmonium-like state $\chi_{c0}(3860)$ was observed in the $e^{+}e^{+}\rightarrow J/\psi D\bar{D}$ process by Belle \cite{Belle:2017egg}. This state has the mass $3862^{+26+40}_{-32-13}$ MeV and the decay width $201^{+154+88}_{-\,\,\,67-82}$ MeV. Its quantum numbers $J^{PC}=0^{++}$ are favored over $2^{++}$ at the level of $2.5\sigma$. In 2019, the narrow resonance state $\psi_3(3842)$ was observed in the decay processes $\psi_3(3842)\to D^0\bar{D}^0$ and $\psi_3(3842)\to D^+D^-$ by LHCb \cite{LHCb:2019lnr}. The measured mass and width are $M=3842.71\pm0.16\pm0.12$ MeV and $\Gamma=2.79\pm0.51\pm0.35$ MeV, respectively. This meson is interpreted as the $1^3D_3$ charmonium state. Later, it was confirmed by BESIII in the $e^+e^-\to\pi^+\pi^-\psi_3(3842)\to\pi^+\pi^-D^+D^-$ process \cite{BESIII:2022quc}. Recently, the BESIII reported the observation of charmonium-like states $\psi(4230)$, $\psi(4500)$, and $\psi(4660)$ in the Born cross sections of process $e^+e^-\to D^{*0}D^{*-}\pi^+$ \cite{BESIII:2023cmv}. The BESIII also studied the $e^+e^-\to D^{*+}_sD^{*-}_s$ process \cite{BESIII:2023wsc} and two significant structures are observed. The first resonance has mass $M=4186.8\pm8.7\pm30$ MeV and width $\Gamma=55\pm15\pm53$ MeV, which is consistent with the $\psi(4160)$ or $\psi(4230)$. The mass and width of the second resonance are measured to be $M=4414.5\pm3.2\pm6.0$ MeV and $\Gamma=122.6\pm7.0\pm8.2$ MeV, respectively, which is consistent with the $\psi(4415)$.

From the theoretical studies, parts of the above mentioned states are good charmonium candidates. However, one still cannot find room to accommodate some charmonium-like states in the conventional quark model picture. They can be alternatively interpreted as tetraquarks, molecules, hybrids, $c\bar{c}$ plus meson-meson continuum components, etc. \cite{Jaffe:2003sg,Swanson:2003tb,Wong:2003xk,Maiani:2004vq,Zhu:2005hp,Suzuki:2005ha,Kalashnikova:2005ui,Kalashnikova:2008qr,Kalashnikova:2009gt,Liu:2008fh,Ebert:2008kb,Liu:2009qhy,Danilkin:2009hr,Danilkin:2010cc,Brambilla:2010cs,Li:2009ad,Coito:2010if,Ferretti:2013faa,Zhou:2013ada,Lu:2016mbb,Qin:2016spb,Lu:2017yhl,Zhou:2017dwj,Guo:2017jvc,Ferretti:2018tco,Cincioglu:2019gzd,Liu:2019zoy,Duan:2020tsx,Duan:2021alw,Dong:2021juy,Wang:2021qus,Peng:2022nrj,Kanwal:2022ani,Chen:2023wfp,TarrusCastella:2024zps}. It is also possible that not all of them exist \cite{Nakamura:2023obk}. For the mysterious state $\chi_{c1}(3872)$ whose width is very narrow, the measured mass is far below the prediction of quenched potential model \cite{Godfrey:1985xj}. The masses of $\psi(4230)$ and $\psi(4360)$ are also inconsistent with the calculated masses with the Godfrey-Isgur (GI) model and a nonrelativistic (NR) model \cite{Barnes:2005pb}. These problems can be partially solved by considering the coupled channel effects \cite{Eichten:2004uh,Kalashnikova:2005ui,Li:2009ad}. In Ref. \cite{Fu:2018yxq}, the authors discussed the coupled channel induced $S$-$D$ mixing effects and calculated the mass spectrum and the open charm decays of $J^{PC}=1^{--}$ charmonium states by using the instantaneous Bethe-Salpeter (BS) equation and the $^3P_0$ model. Their results indicate that the $\psi(4260)$ and $\psi(4360)$ can be assigned as the $4S$-$3D$ mixing charmonium states. The investigations in unquenched quark models of Refs. \cite{Duan:2020tsx,Deng:2023mza} give masses of $\chi_{c0}(2P)$ and $\chi_{c2}(2P)$ consistent with the measured values of $\chi(3915)$ and $\chi(3930)$, respectively. Recently, the strong and electromagnetic decays of the newly observed resonance $\psi_3(3842)$ were studied by using the BS method and the $^3P_0$ model \cite{Li:2023cpl}. The results show that the $\psi_3(3842)$ can be assigned as the charmonium state $\psi_3(1D)$.

To better understand the observed mesons, it is necessary to investigate charmonium states systematically in coupled channel models. One may find such theoretical investigations in the literature, e.g. Refs. \cite{Li:2009ad,Chen:2023wfp,Deng:2023mza}. However, the $S$-$D$ mixing effects are usually ignored for the $J^{PC}=1^{--}$ states. The channel coupling induces not only mass shifts but also the $S$-$D$ mixings for charmonia with quantum numbers $J^{PC}=1^{--}$. In Ref. \cite{Fu:2018yxq}, the authors studied the masses and strong decays of the vector charmonium states by considering the couple-channel and coupled-channel induced $S$-$D$ wave mixing effects. Their results showed that the $2S$-$1D$ states have a small mixing angle and the di-electric decay widths cannot be interpreted well. Since the work focused only on vector mesons, charmonium states with other quantum numbers were not involved. A unified description of all the charmonium states including these effects is still missing. Here, we perform such a study for $c\bar{c}$ states up to the scale around the $D_s^{*+}D_s^{*-}$ threshold.

Besides the spectrum, decay properties of hadrons are also important in understanding their nature. In this work, we will systematically study the masses and open-charm strong decays of various charmonium states in a coupled-channel model using the $^3P_0$ model. The $S$-$D$ mixings will be included for $J^{PC}=1^{--}$ charmonia and the mixing angle will be obtained  by solving the coupled channel model. Based on the extracted mixing angle, we will also calculate the di-electric decay width in order to test our assignments.

This paper is organized as follows. In Sec. \ref{sec2}, we present the framework to investigate the charmonium states. In Sec. \ref{sec3}, we give the numerical results on masses and decay widths. The last section \ref{sec4} is for a short summary and some discussions.

\section{Theoretical Framework}\label{sec2}

\subsection{The quenched potential model}

The quenched potential model does not include the quark-antiquark pair creation and annihilation effects. It usually contains color Coulomb term, linear potential term, and spin-dependent terms. Here, the spin-independent central potential can be written as
\begin{equation}
V_\text{cent}(r)=-\frac{4\alpha_{s}}{3r}+Br+C
\end{equation}
where $B$, $C$, and $\alpha_{s}$ are the string tension, the mass renormalization constant, and the strong coupling constant, respectively. The spin-dependent potential including spin-spin, spin-orbit, and tensor interactions is introduced to describe the fine-structures and hyperfine splittings for the charmonium spectra. It  can be derived from the standard Fermi-Breit expansion to order $\upsilon^2/c^2$. Explicitly, one has
\begin{eqnarray}
V_\text{sd}(r)&=&\frac{32\pi\alpha_{s}}{9m_{c}^2}{\tilde\delta(r)}\mathbf{S_{c} \cdot S_{\bar{c}}}
+\left[\frac{2\alpha_{s}}{m_{c}^2r^3}-\frac{B}{2m_{c}^2r}\right]{\veL
\cdot
\veS}\nonumber \\
&& +\frac{4\alpha_{s}}{3m_{c}^2r^3}\left(\frac{3}{r^{2}}(\mathbf{S_{c}\cdot r})\mathbf{(S_{\bar{c}}\cdot r)}-\mathbf{({S_{c}\cdot S_{\bar{c}}})}\right),
\end{eqnarray}
where $\veS=\mathbf{S_{c}+ S_{\bar{c}}}$ is the total spin of the charm quark and the anticharm quark while $\veL$ is their relative orbital angular momentum. To avoid the failure of nonrelativistic expansion when the two composite quarks are very close to each other, one usually introduces Gaussian smearing for the hyperfine interaction described by the $\delta$ function, $\delta(r)\to{\tilde
\delta}(r)=({\sigma/\pi)^{3}e^{-\sigma^{2}r^{2}}}$ \cite{Barnes:2005pb,Li:2009zu,Lu:2016mbb}. Then, the Hamiltonian in the quenched quark model reads
\begin{equation}
 H_{0}=2m_\text{c}+\frac{\mathbf{p}^2}{m_c}+V_\text{cent}(r)+V_\text{sd}(r),
\end{equation}
with which one obtains the spectra for various charmonium states by solving the Schr$\mathrm{\ddot{o}}$dinger equation.

\renewcommand\tabcolsep{0.13cm}
\renewcommand{\arraystretch}{1.30}


\subsection{The ${^3}{P}{_0}$ model and the coupled-channel model}

The ${^3}{P}{_0}$ model is known as the quark pair creation model in which the $q\bar{q}$ ($q=u,d,s$) pair with $J^{PC}=0^{++}$ is created from vacuum. The model has been widely used to calculate the Okubo-Zweig-Iizuka (OZI) allowed strong decay widths for those states above relevant hadron-hadron thresholds. It was firstly proposed by Micu \cite{Micu:1968mk} and then developed by Le Yaouanc {\it et al} in the late 1970s \cite{LeYaouanc:1972vsx,LeYaouanc:1973ldf}. This model has achieved remarkable success and can be applied to studies of strong decays for the charmonium states \cite{Ortega:2017qmg,Wang:2014lml,He:2014xna,Yang:2009fj}. The model Hamiltonian can be described as
\begin{equation}\label{13}
H_I=2m_q\gamma\int d^3x\bar{\psi}_q\psi_q,
\end{equation}
where $m_q$ is the mass of created quark, $\psi_q$ is the Dirac quark field, and the dimensionless parameter $\gamma$ bespeaks the creation strength for the $q\bar{q}$ pairs from the vacuum.
 We use the effective strength $\gamma_s=\frac{m_{n}}{m_s}\gamma$ for the created strange quark pair. Here, $m_{n}$ ($n=u,d$) and $m_s$ denote the constituent quark masses for the nonstrange and strange quarks, respectively. The $2m_q$ will be canceled by the normalization factor of $\bar{\psi}_q$ and $\psi_q$ in the quenched limit.

The Hamiltonian matrix element of the $A\rightarrow B+C$ process is written as:
\begin{equation}
\left<BC\mid H_I\mid A\right>=\delta^3(\mathbf{P}_{A}-\mathbf{P}_{B}-\mathbf{P}_{C}){\cal M}^{M_{J_{A}}M_{J_{B}}M_{J_{C}}},
\end{equation}
where the Hamiltonian operator $H_I$ is
\begin{eqnarray}
 H_I&=&-3\gamma\sum_{m} \left<1m1-m\mid00\right>\int \limits d\mathbf{p}_3d\mathbf{p}_4\delta^3(\mathbf{p}_3+\mathbf{p}_4)\nonumber\\
 &&\times{\cal Y}_{1}^{m}\left(\frac{\mathbf{p}_3-\mathbf{p}_4}{2}\right)\chi_{1-m}^{34}\varphi_{0}^{34}\omega_{0}^{34}b_{3}^{\dagger}(\mathbf{p}_3)d_{4}^{\dagger}(\mathbf{p}_4).
\end{eqnarray}
The numbers 3 and 4 in the superscripts/subscripts label the created quark and antiquark, respectively. Their corresponding momenta are $\mathbf{p}_3$ and $\mathbf{p}_4$. $\varphi_{0}^{34}$ and $\omega_{0}^{34}$ indicate singlet wavefunctions of the $q\bar{q}$ pair in the flavor and color spaces, respectively. Since the quantum numbers $J^{PC}$ for the created $q\bar{q}$ pair must be $0^{++}$, its spin wave function is denoted as $\chi_{1-m}^{34}$ and the spatial relative motion is described by ${\cal Y}_{1}^{m}(\mathbf{p})=\mid p\mid Y_{1}^{m}(\theta_{p}, \phi_{p})$. $b_{3}^{\dagger}(\mathbf{p}_3)$ and $d_{4}^{\dagger}(\mathbf{p}_4)$ denote the creation operators of quark and antiquark, respectively. In this model, the helicity amplitude is expressed as
\begin{eqnarray}
&&\mathcal{M}^{M_{J_A}M_{J_B} M_{J_C}}(\mathbf{P})\nonumber\\
&=&\gamma\sqrt{96\pi} \sum_{M_{L_A},M_{L_B},M_{L_C}}\sum_{M_{S_A},M_{S_B},M_{S_C},m}
I^{M_{L_A},m}_{M_{L_B},M_{L_C}}({\textbf{P}})\nonumber\\
 && \nonumber   \times\langle
L_A M_{L_A}; S_A M_{S_A} | J_A M_{J_A} \rangle\langle 1\;m;1\;-m|\;0\;0 \rangle\nonumber\\
&&
\times\langle L_B M_{L_B};
S_B M_{S_B} | J_B M_{J_B} \rangle\langle
L_C M_{L_C}; S_C M_{S_C} | J_C M_{J_C} \rangle \nonumber\\
&&
\times\langle\phi^{1 3}_B \phi^{2 4}_C | \phi^{1 2}_A \phi^{3 4}_0\rangle
\langle \chi^{1 3}_{S_B M_{S_B}}\chi^{2 4}_{S_C M_{S_C}}  | \chi^{1 2}_{S_A M_{S_A}} \chi^{3 4}_{1 -\!m} \rangle\;.
\end{eqnarray}
Here, $I^{M_{L_A},m}_{M_{L_B},M_{L_C}}({\textbf{P}})$ is the momentum space integral of the simple harmonic oscillator $({\textbf{SHO}})$ wave functions,
\begin{align}
&I^{M_{L_A},m}_{M_{L_B},M_{L_C}}({\textbf{P}})=\int d^3\mathbf{p}_3 \Psi^*_{n_B L_B M_{L_B}}\left(\frac{m_3 \mathbf{P}}{m_1+m_3}+\mathbf{p}_3 \right) 
\nonumber\\
&\times\Psi^*_{n_C L_C M_{L_C}}\left(\frac{m_3 \mathbf{P}}{m_2+m_3}+\mathbf{p}_3 \right)\Psi_{n_A L_A M_{L_A}}\left(\mathbf{P}+\mathbf{p}_3 \right)\mathcal{Y}_{1m}(\mathbf{p}_3).
\end{align}
In this equation, the momenta of final state mesons satisfy the relation $\mathbf{P}=\mathbf{P}_B=-\mathbf{P}_C$ in the center-of-mass frame and $\mathbf{p}=\mathbf{p}_3$ is the momentum of the created quark.
The SHO wave function reads
\begin{eqnarray}\label{wave function}
	\Psi_{n L M_{L}}(\mathbf{P})&=\frac{(-1)^n(-i)^L}{\beta^{3/2}}\sqrt{\frac{2n!}{\Gamma(n+L+3/2)}}\left(\frac{\mathbf{P}}{\beta}\right)^L e^-{\frac{\mathbf{P}^2}{2\beta^2}}\nonumber\\
	&\times
	L^{L+1/2}\left(\frac{\mathbf{P^2}}{\beta^2}\right)Y_{LM_{L}}(\Omega_\mathbf{P}),
\end{eqnarray}
where $\beta$ is the harmonic oscillator parameter.
 To compare with the experimental data, we should convert the helicity amplitude $\mathcal{M}^{M_{J_A}M_{J_B} M_{J_C}}(\mathbf{P})$ to the partial wave amplitude $\mathcal{M}^{JL} (\mathbf{P})$
\begin{eqnarray}\label{integral}
&\mathcal{M}^{JL} (A\rightarrow B+C)=\frac{\sqrt{4\pi(2 L+1)}}{2 J_A+1} \!\! \sum_{M_{J_B},M_{J_C}} \langle L 0 J M_{J_A}|J_A M_{J_A}\rangle \nonumber\\
&\times  \langle J_B M_{J_B} J_C M_{J_C} | J M_{J_A} \rangle \mathcal{M}^{M_{J_A} M_{J_B} M_{J_C}}({\textbf{P}}).
\end{eqnarray}

For higher chanmonium states lying above their respective thresholds of open-charm decay channels, their mass spectrum can be strongly affected by the coupled-channel effects. Incorporating such effects in the quenched potential model, one refers to it as the coupled-channel model. In this coupled-channel model, a quarkonium state couples to the continuum states, which modifies the mass of that near-threshold charmonium.

The dominant decay mode of a charmonium involves the open-charm meson-antimeson pair where the open-charm meson $c\bar{q}$ means $S$-wave $D$, $D^\ast$, $D_{s}$, or $D_{s}^\ast$ and its antiparticle $\bar{c}q$ is simply denoted as $\bar{D}^{(*)}$ or $\bar{D}_s^{(*)}$. In the coupled-channel model, the wave function of a physical state can be written as
\begin{equation}
\mid\Psi\rangle=\left[ C_{A}|\psi
_{A}\rangle\atop \sum\limits_{BC}\int C_{BC}(\mathbf{P})
|BC,\mathbf{P}\rangle d^3\mathbf{P}\right],
\end{equation}
where $|\psi_{A}\rangle$ is the bare charmonium state and $|BC,\mathbf{P}\rangle$ represents an open-charm meson-antimeson continuum state with $\mathbf{P}$ being the momentum of the open-charm meson in the center-of-mass frame. The factors $C_{A}$ and $C_{BC}(\mathbf{P})$ denote the normalization constants of the bare state and the meson continuum state, respectively. The normalization condition reads 

 \begin{equation}\label{normalization}
	\mid C_A\mid^2+\sum_{BC}\int\mid C_{BC}(\mathbf{P})\mid^2 d^3\mathbf{P}= 1.
\end{equation}
Then, the Hamiltonian $H$ of the coupled-channel model has the form
\begin{equation}
H= \left[\begin{array}{cc}
H_A&H_I\\
H_I&H_{BC}
\end{array}\right]
\end{equation}
and the physical mass $M$ satisfies the Schrodinger equation
\begin{equation}\label{phy mass}
H |\Psi\rangle = M |\Psi\rangle.
\end{equation}
In the Hamiltoinan, $H_A$ acting only on the initial wave function $\mid\psi_A\rangle$ is the part for the initial charmonium state. The bare charmonium mass $M_A$ is obtained by solving
\begin{equation}\label{bare mass}
H_A\mid\psi_A\rangle=M_A\mid\psi_A\rangle.
\end{equation}
$H_{BC}$ denotes the Hamiltonian for the free meson-antimeson pair, which means that the interactions between $B$ and $C$ is neglected and thus the energy of meson continuum can be expressed as $E_{BC}=\sqrt{m^2_B+\mathbf{P}^2}+\sqrt{m^2_C+\mathbf{P}^2}$. One has
\begin{equation}\label{EBC}
H_{BC}\mid BC, \mathbf{P}\rangle=E_{BC}\mid BC, \mathbf{P}\rangle.
\end{equation}
The $H_I$ part given in Eq. \eqref{13} describes the interactions between the bare charmonium and the final open-charm meson-antimeson states.

From the matrix form of the coupled-channel Schrodinger equation
\begin{equation}
\begin{split}
\left[\begin{array}{cc}
H_A&H_I\\
H_I&H_{BC}
\end{array}\right]
\left[C_{A}|\psi
_{A}\rangle\atop \sum\limits_{BC}\int C_{BC}(\mathbf{P})
|BC,\mathbf{p}\rangle d^3\mathbf{P}\right]\\
=M\left[C_{A}|\psi
_{A}\rangle\atop \sum\limits_{BC}\int C_{BC}(\mathbf{P})
|BC,\mathbf{\mathbf{P}}\rangle d^3\mathbf{P}\right].
\end{split}
\end{equation}
For a charmonium state below relevant thresholds, one obtains
\begin{equation}\label{M&MA}
	M-M_A-\Pi_{BC}(M)=0.
\end{equation}
Here, the self-energy function $\Pi_{BC}$ is explicitly given by
\begin{eqnarray}\label{PiBC}
	\Pi_{BC}(M)
	&=&\sum_{BC}\int_0^\infty\frac{1}{M-E_B-E_C}\sum_{JL}\mid\mathcal{M}_{JL}\mid^2 P^2dP\nonumber\\,
\end{eqnarray}
where the sum of $BC$ covers the meson-antimeson pairs $D\bar{D}$, $D\bar{D}^*$, $D^*\bar{D}^*$, $D_s\bar{D}_s$, $D_s\bar{D}^*_s$, and $D^*_s\bar{D}^*_s$\footnote{For convenience, we will always use $D\bar{D}^*$ to denote the open-charm meson-antimeson pair $D\bar{D}^*/D^*\bar{D}$. Similarly, $D_s\bar{D}_s^*$ denotes $D_s^+D_s^{*-}/D_s^{*+}D_s^-$.}. Each meson-antimeson channel may contribute to the mass shift of the charmonium.
 In general, we should consider all charmed meson loops. However, including an unlimited number of loops leads to  difficulties in calculating the self-energy function \cite{Deng:2023mza}. Determining the optimal number of mass thresholds is challenging. A commonly used approach is to concentrate only on the S-wave charmed mesons and ignore the rest \cite{Li:2009ad,Fu:2018yxq}. 
In this case, there is no singularity problem and the self-energy function denotes the sum of the mass shifts.

According to the normalization condition Eq. (\ref{normalization}), 
the proportion of the charmonium in the physical state is  given by \cite{Liu:2011yp}
\begin{equation}\label{component}
	P_{c\bar{c}}=\left[1+\sum_{BC}\int_0^\infty\frac{1}{(M-E_B-E_C)^2}\sum_{JL}\mid\mathcal{M}_{JL}\mid^2P^2dP\right]^{-1}.
\end{equation}

 For a charmonium state above some threshold(s), Eq. \eqref{M&MA} can be analytically continued to the complex plane, allowing for complex number solutions. However, if we assume that $\Gamma_{\text{total}}\ll M$, the real pole mass $M$ can be defined as 
\begin{equation}\label{ReM&MA}
M-M_A-\text{Re}[\Pi_{BC}(M)]=0,
\end{equation}
   and the decay width as
\begin{equation}\label{two body decay}
	\Gamma_\text{total}=-2\text{Im}(\Pi_{BC}(M)).
\end{equation}
In this case, the singularity occurs in Eq. \eqref{ReM&MA}. We introduce the principal part integral $\mathcal{P}$ to solve this problem.
The real value of the self-energy function $\Pi_{BC}$ is explicitly given by
\begin{eqnarray}\label{PiBC}
\text{Re}[\Pi_{BC}(M)]
&=&\sum_{BC}\mathcal{P}\int_0^\infty\frac{1}{M-E_B-E_C}\sum_{JL}\mid\mathcal{M}_{JL}\mid^2 P^2dP.\nonumber\\
\end{eqnarray}   
The imaginary part $\text{Im}(\Pi_{BC}(M))$ can be related to the amplitude of the $^3P_0$ model 
\begin{eqnarray}\label{Im}
	\text{Im}[\Pi_{BC}(M)]=-\sum_{BC}\pi \frac{\mathbf{P}E_BE_C}{M}\sum_{JL}\mid\mathcal{M}_{JL}\mid^2,
\end{eqnarray}
where $\mathbf{P}$ is the momentum of the final state mesons. 
The real pole mass $M$ can be calculated by solving  Eq. \eqref{ReM&MA}.
Then, we apply this value to get the width $\Gamma_{\text{total}}$.

It should be noted that  $P_{c\bar{c}}$  becomes a complex number for a resonance.
Therefore, the probability of the $c\bar{c}$ component cannot be strictly defined. 
Nevertheless, the authors of Refs. \cite{Guo:2015daa,Sekihara:2014kya} obtain this probability  by taking some approximations into account .
Here, we utilize the assump tion that the resonance states have small decay widths.
Based on this assumption, the proportion of the charmonium can be crudely calculated using Eq. \eqref{component}, which helps to some extent in defining the $c\bar{c}$ component.
 The above discussions are based on the adopted assumption that  the charmonium state is a narrow resonance \cite{Zhou:2017dwj,Duan:2020tsx}. 
 Although this method is crude, it helps in understanding the basic features of mass shifts, hadron masses, and the probability of the $c\bar{c}$ component. 
For the coupled-channel model parameters, we refit the measured masses of the charmonium states. The following values will be employed, $m_n$=0.33 GeV, $m_s$=0.55 GeV, $m_{c}=1.700$ GeV, $B=0.13$ GeV$^2$, $ \alpha_{s}=0.621\notag$, $\sigma=1.802$ GeV, $C =-0.233$ GeV, $\gamma=0.20$, and $\beta=0.31$ GeV.

The $^3S_1$ and $^3D_1$ $c\bar{c}$ states have the same quantum numbers $J^{PC}=1^{--}$. The calculated di-electric decay widths of the pure $nD$ states are highly suppressed compared to the experimental values. In Refs. \cite{Kuang:1989ub,Ding:1991vu,Rosner:2001nm,Badalian:2017nyv}, the authors solved this issue by considering the $S$-$D$ wave mixing effect. The mixing angle is extracted from the ratio of di-electric widths between the observed states that are mixtures of $S$- and $D$-wave charmonia,
\begin{eqnarray}\label{mixtureangle}
\tilde{\psi}((n+1)S) &=& |(n+1)S\rangle \mathrm{cos\theta}+|nD\rangle \mathrm{sin\theta},\nonumber\\
\tilde{\psi}(nD) &=& -|(n+1)S\rangle \mathrm{sin\theta}+|nD\rangle \mathrm{cos\theta}.
\end{eqnarray}
In Refs. \cite{Heikkila:1983wd,Lu:2016mbb,Fu:2018yxq}, the authors also discussed the $S$-$D$ mixing for vector quarkonium states. In their scheme, this mixing is caused by the couple channel effects and the mixing angle can be determined from the coupled channel equations. In the present work, we follow this idea and make an extension for the study of \cite{Fu:2018yxq}. Assuming that the contribution of meson continuum has been subtracted from the physical states, one may focus only on the role of $S$-$D$ mixing. The physical states are also written as those in Eq. \eqref{mixtureangle}. The $S$-$D$ wave mixing angle can be obtained by solving the reexpressed equation of Eq. \eqref{ReM&MA},
\begin{equation}\label{mix}
\text{det} \left|\begin{array}{cc}
M-M_S-\text{Re}[\Pi_{SS}(M)]&\text{Re}[\Pi_{SD}(M)]\\
\text{Re}[\Pi_{DS}(M)]&M-M_D-\text{Re}[\Pi_{DD}(M)]
\end{array}\right|=0,
\end{equation}
where $\text{Re}[\Pi_{SS}(M)]$ and $\text{Re}[\Pi_{DD}(M)]$ are described by Eq. \eqref{PiBC} and
\begin{equation}
\text{Re}[\Pi_{SD}(M)]=\sum_{BC}\int_0^\infty\frac{\langle \psi_S\mid H_I\mid BC,P\rangle\langle BC,P\mid H_I\mid \psi_D\rangle}{M-E_B-E_C-i\varepsilon}P^2dP.
\end{equation}

\subsection{The di-electric decay  width }

The di-electric decay width of a vector charmonium state including radiative QCD corrections \cite{Kwong:1987ak} can be expressed as
 \begin{equation}
\Gamma_{ee}((n+1)S)=\frac{4\alpha^2e_c^2\mid R_{(n+1)S}(0)\mid^2}{M_{(n+1)S}^2}\left(1-\frac{16\alpha_s}{3\pi}\right)
\end{equation}
 \begin{equation}
\Gamma_{ee}(nD)=\frac{25\alpha^2e_c^2\mid R^{\prime\prime}_{nD}(0)\mid^2}{2M_{nD}^2M_c^4}\left(1-\frac{16\alpha_s}{3\pi}\right),
\end{equation}
where $e_c=2/3$ is the charm quark charge in the unit of the electron charge and $\alpha=1/137$ is the fine-structure constant. $R_{nS}(0)$ and $R^{\prime\prime}_{nD}(0)$ denote the radical $S$-wave function and the second derivative of the radical $D$-wave function at the origin, respectively. $M_{(n+1)S}$ is the mass of the $(n+1)^3S_1$ charmonium state and $ M_{nD}$ is that of the $n^3D_1$ state. The factor $(1-16\alpha_s/3\pi)$ is from the QCD radiative correction.

With the assumption that the $1^{--}$ charmonia are mixtures of $(n+1)^3S_1$ and $(n)^3D_1$ states, the expressions of their di-electric decay width can be written as
\begin{eqnarray}
\Gamma_{ee}((n+1)S)&=&\frac{4\alpha^2e_c^2}{M_{(n+1)S}^2} \left| \mathrm{sin\theta} R_{(n+1)S}(0)+  \frac{5\mathrm{cos\theta}}{2\sqrt{2}m_b^2}R^{\prime\prime}_{nD}(0)\right|^2\nonumber\\
&&\times\left(1-\frac{16\alpha_s}{3\pi}\right),\\
\Gamma_{ee}(nD)&=&\frac{4\alpha^2e_c^2}{M_{nD}^2}\left| \mathrm{cos\theta} R_{(n+1)S}(0)- \frac{5\mathrm{sin\theta}}{2\sqrt{2}m_b^2}R^{\prime\prime}_{nD}(0)\right|^2\nonumber\\
&&\times\left(1-\frac{16\alpha_s}{3\pi}\right),
\end{eqnarray}
where $\theta$ is extracted by solving Eq. \eqref{mix}.

\section{Numerical results}\label{sec3}

\renewcommand\tabcolsep{0.11cm}
\renewcommand{\arraystretch}{1.05}

\begin{table*}[!htbp]
\caption{\label{mass}Masses of charmonium states obtained in the adopted coupled-channel model with $\gamma=0.2$ and $\beta=0.31$ GeV. The contributions to mass shifts from the six open-charm channels are shown in 2nd$-$7th columns. The total mass shift $\delta m$ is the summation of these contributions. The mass $M_A$ is the bare mass of the charmonium. The $M_{\rm cou}=M_A+\delta m$ and $M_{\rm non}$ denote the calculated masses in the coupled-channel model we use and in a nonrelativistic quark potential model adopted in Ref. \cite{Barnes:2005pb}, respectively. The experimental data $M_{\rm exp}$ are taken from the PDG particle data book \cite{ParticleDataGroup:2022pth}. The $P_{c\bar{c}}$ is the probability of the $c\bar{c}$ component in the charmonium. The number $0.0$ indicates that the corresponding open-charm channel does not induce mass shift, while the symbol ``$-$'' means that there is no available experimental datum or no theoretical result in Ref. \cite{Barnes:2005pb}. All the masses are given in units of MeV.}
\scalebox{1}[1]{
\begin{tabular*}{2.0\columnwidth}{@{\extracolsep{\fill}}c c c c c c c c c c c c c@{}}
\toprule[1pt]\toprule[1pt]
Meson & $D\bar{D}$ & $D\bar{D}^{*}$ & $D^*\bar{D}^*$ &$D_s\bar{D}_s$  &$D_s\bar{D}^*_s$ &$D^*_s\bar{D}^*_s$&$\delta m$&$M_A$&$M_\text{cou}$&$M_\text{non}$&$M_\text{exp}$&$P_{c\bar{c}}(\%)$  \\\hline

$\eta_c(1S)(1{^1}S_0)$ &$0.0$&$-15.8$&$-14.1$&$0.0$&$-4.9$&$-4.5$&$-39.3$&3018.3&2979.1&2982&$2983.9\pm0.4$&96.9\\
$J/\psi(1S)(1{^3}S_1)$ &$-3.4$&$-11.7$&$-18.1$&$-1.0$&$-3.6$&$-5.7$&$-43.4$&3136.1&3092.7&3090&$3096.900\pm0.006$&96.2\\
$\eta_c(2S)(2{^1}S_0)$ &$0.0$&$-22.5$&$-18.2$&$0.0$&$-6.3$&$-5.5$&$-52.4$&3692.4&3639.9&3630&$3637.5\pm1.1$&92.5\\
$\psi(2S)(2{^3}S_1)$  &$-5.9$&$-16.0$&$-22.3$&$-1.3$&$-4.3$&$-6.6$&$-56.4$&3728.9&3672.5&3672&$3686.10\pm0.06$&90.3\\
$\eta_c(3S)(3{^1}S_0)$ &$0.0$&$-28.5$&$-25.1$&$0.0$&$-6.9$&$-5.9$&$-66.4$&4083.2&4016.8&4043&$-$&61.2\\
$\psi(4040)(3{^3}S_1)$ &$-2.3$&$-20.2$&$-27.2$&$-1.3$&$-4.9$&$-7.1$&$-62.9$&4107.3&4044.4&4072&$4039\pm1$&72.1\\
$\eta_c(4S)(4{^1}S_0)$ &$0.0$&$-16.8$&$-23.0$&$0.0$&$-7.2$&$-6.5$&$-53.5$&4398.8&4345.3&4384&$-$&95.5\\
$\psi(4360)(4{^3}S_1)$ &$-4.6$&$-11.1$&$-26.5$&$-1.4$&$-5.0$&$-7.6$&$-56.3$&4417.5&4361.2&4406&$4372\pm9$&92.7\\
$h_{c}(1P)(1{^1}P_1)$ &$0.0$&$-39.3$&$-20.9$&$0.0$&$-6.6$&$-5.7$&$-72.5$&3596.0&3523.5&3516&$3525.38\pm0.11$&90.2\\
$\chi_{c0}(1P)(1{^3}P_0)$ &$-9.2$&$0.0$&$-31.3$&$-1.8$&$0.0$&$-9.1$&$-51.4$&3479.2&3427.8&3424&$3414.71\pm0.30$&93.5\\
$\chi_{c1}(1P)(1{^3}P_1)$ &$0.0$&$-33.4$&$-24.0$&$0.0$&$-4.9$&$-6.9$&$-69.2$&3567.5&3498.3&3505&$3510.67\pm0.05$&91.0\\
$\chi_{c2}(1P)(1{^3}P_2)$ &$-8.5$&$-19.4$&$-23.5$&$-2.1$&$-5.2$&$-6.0$&$-64.6$&3635.6&3570.9&3556&$3556.17\pm0.07$&90.4\\
$h_{c}(2P)(2{^1}P_1)$ &$0.0$&$-54.4$&$-22.0$&$0.0$&$-6.9$&$-5.8$&$-89.0$&3989.8&3900.7&3934&$-$&73.8\\
$\chi_{c0}(3860)(2{^3}P_0)$ &$-16.6$&$0.0$&$-33.2$&$-2.7$&$0.0$&$-9.4$&$-61.8$&3920.5&3858.7&3852&$3862^{+26+40}_{-32-13}$&89.0\\
$\chi_{c1}(3872)(2{^3}P_1)$ &$0.0$&$-57.5$&$-23.6$&$0.0$&$-5.4$&$-6.7$&$-93.2$&3963.3&3870.1&3925&$3871.65\pm0.06$&43.4\\
$\chi_{c2}(3930)(2{^3}P_2)$ &$-4.4$&$-24.6$&$-28.0$&$-2.5$&$-5.3$&$-6.5$&$-71.2$&4024.5&3953.3&3972&$3922.5\pm1.0$&75.1\\
$h_{c}(3P)(3{^1}P_1)$ &$0.0$&$-28.3$&$-25.3$&$0.0$&$-7.0$&$-6.4$&$-67.0$&4309.4&4242.4&4279&$-$&87.1\\
$\chi_{c0}(3P)(3{^3}P_0)$ &$-1.1$&$0.0$&$-38.1$&$-3.1$&$0.0$&$-10.2$&$-52.5$&4257.6&4205.2&4202&$-$&87.1\\
$\chi_{c1}(3P)(3{^3}P_1)$ &$0.0$&$-30.1$&$-27.1$&$0.0$&$-5.5$&$-7.2$&$-69.9$&4283.0&4213.0&4271&$-$&78.0\\
$\chi_{c2}(3P)(3{^3}P_2)$ &$-9.0$&$-13.9$&$-28.4$&$-1.9$&$-5.5$&$-7.2$&$-65.9$&4341.6&4275.8&4317&$-$&85.6\\
$\eta_{c}(1D)(1{^1}D_2)$ &$0.0$&$-52.2$&$-24.1$&$0.0$&$-6.8$&$-5.7$&$-88.9$&3873.5&3784.6&3799&$-$&80.6\\
$\psi(3770)(1{^3}D_1)$ &$-23.9$&$-7.4$&$-36.8$&$-1.4$&$-1.0$&$-9.5$&$-80.1$&3843.2&3763.2&3785&$3773.7\pm0.4$&81.7\\
$\psi_{2}(3823)(1{^3}D_2)$ &$0.0$&$-49.1$&$-27.1$&$0.0$&$-5.6$&$-6.8$&$-88.5$&3872.2&3783.7&3800&$3823.5\pm0.5$&80.2\\
$\psi_{3}(3842)(1{^3}D_3)$ &$-15.2$&$-23.9$&$-25.6$&$-2.6$&$-5.5$&$-5.4$&$-78.1$&3883.8&3805.6&3806&$3842.71\pm0.20$&82.0\\
$\eta_c(2D)(2{^1}D_2)$ &$0.0$&$-27.5$&$-26.2$&$0.0$&$-7.6$&$-6.1$&$-67.3$&4203.5&4136.2&4158&$-$&89.9\\
$\psi(4160)(2{^3}D_1)$ &$3.0$&$-5.6$&$-42.1$&$-1.8$&$-1.5$&$-9.7$&$-57.7$&4169.3&4111.6&4142&$4191\pm5$&87.9\\
$\psi_{2}(2D)(2{^3}D_2)$ &$0.0$&$-24.1$&$-29.3$&$0.0$&$-6.1$&$-7.1$&$-66.6$&4202.2&4135.6&4158&$-$&84.3\\
$\psi_{3}(2D)(2{^3}D_3)$ &$-12.5$&$-16.5$&$-28.0$&$-2.1$&$-6.1$&$-6.0$&$-71.2$&4214.7&4143.5&4167&$-$&76.4\\
$\eta_{c}(3D)(3{^1}D_2)$ &$0.0$&$-39.0$&$-19.4$&$0.0$&$-7.5$&$-6.1$&$-72.0$&4491.0&4418.9&$-$&$-$&84.8\\
$\psi(4415)(3{^3}D_1)$ &$-1.4$&$-0.3$&$-30.5$&$-1.5$&$-1.5$&$-9.7$&$-45.0$&4453.4&4408.4&$-$&$4421\pm4$&93.4\\
$\psi_{2}(3D)(3{^3}D_2)$ &$0.0$&$-31.9$&$-21.8$&$0.0$&$-6.6$&$-7.0$&$-67.3$&4489.7&4422.3&$-$&$-$&89.7\\
$\psi_{3}(3D)(3{^3}D_3)$ &$-12.1$&$-23.3$&$-18.5$&$-2.2$&$-5.5$&$-6.1$&$-67.8$&4502.9&4435.2&$-$&$-$&82.9\\
$h_{c}(1F)(1{^1}F_3)$ &$0.0$&$-72.1$&$-26.6$&$0.0$&$-6.5$&$-5.2$&$-110.5$&4086.7&3976.2&4026&$-$&86.9\\
$\chi_{c2}(1F)(1{^3}F_2)$ &$7.3$&$-3.7$&$-47.1$&$-1.9$&$-1.4$&$-9.7$&$-56.6$&4082.6&4026.0&4029&$-$&71.0\\
$\chi_{c3}(1F)(1{^3}F_3)$ &$0.0$&$-69.2$&$-28.8$&$0.0$&$-5.6$&$-6.1$&$-109.7$&4088.1&3978.4&4029&$-$&85.6\\
$\chi_{c4}(1F)(1{^3}F_4)$ &$-15.3$&$-30.9$&$-30.8$&$-2.9$&$-5.4$&$-4.7$&$-90.0$&4087.3&3997.3&4021&$-$&75.2\\
\bottomrule[1pt]\bottomrule[1pt]
\end{tabular*}}
\end{table*}

\begin{figure*}
\includegraphics[width=500pt]{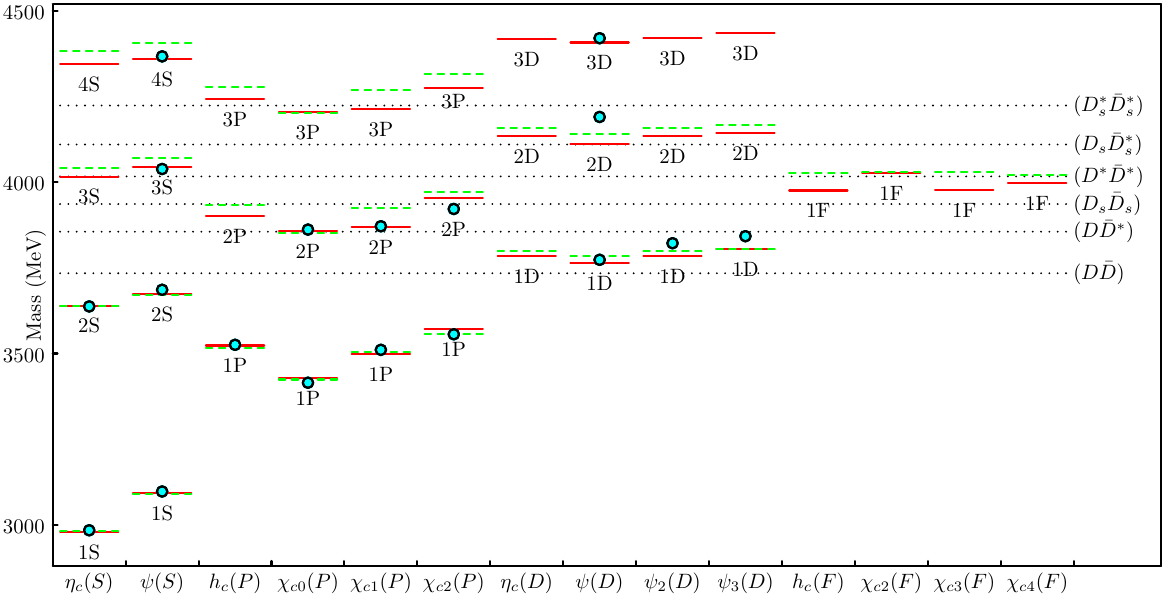}
\caption{Relative positions for the charmonium states. The red solid and green dashed lines denote the calculated masses (table \ref{mass} and table \ref{massmix}) in our coupled-channel model and those in the nonrelativistic potential model of Ref. \cite{Barnes:2005pb}, respectively. We use cyan dots to denote the central values of experimental results \cite{ParticleDataGroup:2022pth}. The blue dotted lines are related open-charm meson-antimeson thresholds.} \label{charmonium}
\end{figure*}

\begin{table*}
\caption{\label{S-wavedecay}Two-body open-charm strong decay widths of different channels for $3S$ and $4S$ charmonium states in units of MeV. $\Gamma_{\rm th}$ and $\Gamma_{\rm exp}$ denote our results and experimental measurements, respectively. The values in the square brackets mean the total widths. When the experimental datum is not available, a symbol ``$-$'' is presented.}
	\begin{tabular}{c  c  l  c  c  c  c  l  c  c }\toprule[1pt]\toprule[1pt]
		Meson &State&Channel& $\Gamma_\text{th}$&$\Gamma_\text{exp}$   &Meson &State&Channel& $\Gamma_\text{th}$&$\Gamma_\text{exp}$\\
		\hline
		$\eta_{c}(3S)$&   $3{^1}S_0$ &   $D\bar{D}^{*}$ & $3.7$&      &$ \psi(4040)$
&                                                                 $3{^3}S_1$&   $D\bar{D}$&     $5.1$&        \\
            &             &   $D^*\bar{D}^*$ &  $1.4$ &    &  &    &$D\bar{D}^{*}$ &           $8.8$& \\
            &             &                      &    &  &    &&$D^*\bar{D}^*$ &           $22.3$&                \\
            &             &                      &    &  &  & &$D_s\bar{D}_s$ &           $0.1$&                  \\
            &             & &$[5.1]$&$[-]$& &  &  &$[36.3]$& $[80\pm10]$            \\

$\eta_{c}(4S)$&     $4{^1}S_0$&   $D\bar{D}^*$&    $17.6$&     &$\psi(4360)$&     $4{^3}S_1$&    $D\bar{D}$&       $0.2$&       \\
            &  &  $D^*\bar{D}^*$&     $6.8$      &&&&                                         $D\bar{D}^*$&           $11.1$&       \\
            &  &  $D_s\bar{D}^*_s$&   $0.1$      &&&&                                         $D^*\bar{D}^*$&         $10.9$&       \\
            &  &  $D^*_s\bar{D}^*_s$& $0.2$      &&&&                                         $D_s\bar{D}_s$&         $0.3 $&      \\
            &  &  $D\bar{D}^*_0$&     $0.0$      &&&&                                         $D^*_s\bar{D}^*_s$&     $0.4$      \\
            &  &  $D\bar{D}_1$&       $0.0$        &&&&                                         $D\bar{D}_1$&           $6.9$     \\
            &  &  $D\bar{D}^\prime_1$&$0.0$        &&&&                                         $D\bar{D}^\prime_1$ &   $1.7$       \\
            &  &  $D\bar{D}^*_2$ &    $4.4$       &&&&                                         $D\bar{D}^*_2$ &        $5.6$       \\
            &  &  $D^*\bar{D}^*_0$ &  $0.0$        &&&&                                         $D^*\bar{D}^*_0$ &      $0.4$    \\

             &             &&$[29.1]$&$[-]$ & &  & &$[37.5]$& $[115\pm13]$                       \\
 \bottomrule[1pt]\bottomrule[1pt]
	\end{tabular}
\end{table*}

\begin{table*}[!htbp]
	\caption{\label{D-wavedecay}Two-body open-charm strong decay widths of different channels for $1D$, $2D$, and $3D$ charmonium states in units of MeV. $\Gamma_{\rm th}$ and $\Gamma_{\rm exp}$ denote our results and experimental measurements, respectively. The values in the square brackets mean the total widths. When the experimental datum is not available, a symbol ``$-$'' is presented.}
\begin{tabular}{c  c  l  c  c  c  c  l  c  c }\toprule[1pt]\toprule[1pt]
		Meson &State&Channel& $\Gamma_\text{th}$&$\Gamma_\text{exp}$   &Meson &State&Channel& $\Gamma_\text{th}$&$\Gamma_\text{exp}$\\\hline
	$\psi(3770)$&
	$1{^3}D_1$ &   $D^+D^-$ &       $9.4$&         &$\psi_{3}(3842)$&   $1{^3}D_3$     &$D^+D^-$      &        $0.8$&  \\
		&&       $D^0\bar{D}^0$&  $13.9$   &&&                                        &$D^0\bar{D}^0$&        $1.1$&  \\
		&             & &$[23.3]$&$[27.2\pm1.0]$ &  &        & &$[1.9]$& $[2.8\pm0.6]$          \\
		$\psi(4160)$ &$2{^3}D_1$&$D\bar{D}$&13.4&&$\psi_2(2D)$&$2{^3}D_2$&$D\bar{D}$&  	0.0&\\
		&&$D\bar{D}^*$&14.2&&&&$D\bar{D}^*$&34.7&\\
		&&$D^*\bar{D}^*$&19.5&&&&$D^*\bar{D}^*$&18.0&\\
		&&$D_s\bar{D}_s$&0.2&&&&$D_s\bar{D}_s$&  	0.0&\\
		&&$D_s\bar{D}^*_s$&1.2&&&&$D_s\bar{D}^*_s$&2.7&\\
		&&&[48.6]&$[70\pm10]$&&&&[55.4]&$[-]$\\
		$\psi_3(2D)$ &$2{^3}D_3$&$D\bar{D}$&2.7&&$\eta_c(2D)$&$2{^1}D_2$&$D\bar{D}$&0.0&\\
		&&$D\bar{D}^*$&5.3&&&&$D\bar{D}^*$&26.8&\\
		&&$D^*\bar{D}^*$&10.4&&&&$D^*\bar{D}^*$&13.6&\\
		&&$D_s\bar{D}_s$&1.3&&&&$D_s\bar{D}^*_s$&2.1&\\
		&&$D_s\bar{D}^*_s$&0.5&&&&&&\\
		&&&[20.3]&$[-]$&&&&[42.5]&$[-]$\\
		$\psi(4415)$ &$3{^3}D_1$&$D\bar{D}$&2.1&&$\psi_2(3D)$&$3{^3}D_2$&$D\bar{D}$&0.0&\\
		&&$D\bar{D}^*$&8.4&&&&$D\bar{D}^*$&14.3&\\
		&&$D^*\bar{D}^*$&16.7&&&&$D^*\bar{D}^*$&13.2&\\
		&&$D_s\bar{D}_s$&0.9&&&&$D_s\bar{D}_s$&0.0&\\
		&&$D_s\bar{D}^*_s$&0.1&&&&$D_s\bar{D}^*_s$&0.6&\\
		&&$D^*_s\bar{D}^*_s$&0.5&&&&$D^*_s\bar{D}^*_s$&0.3&\\
		&&$D\bar{D}_1$&6.8&&&&$D\bar{D}_1$&0.0&\\
		&&$D\bar{D}^{\prime}_1$&1.1&&&&$D\bar{D}^{\prime}_1$&0.0&\\
		&&$D\bar{D}^*_0$& 0.0&&&&$D\bar{D}^*_0$&0.1&\\
		&&$D\bar{D}^*_2$&3.3&&&&$D\bar{D}^*_2$&9.1&\\
		&&$D^*\bar{D}^*_0$&2.7&&&&$D^*\bar{D}^*_0$&1.4&\\
		&&&[42.4]&$[62\pm20]$&&&&[38.9]&$[-]$\\
		$\psi_3(3D)$ &$3{^3}D_3$&$D\bar{D}$&5.7&&$\eta_{c}(3D)$&$3{^1}D_2$&$D\bar{D}$&0.0&\\
		&&$D\bar{D}^*$&0.0&&&&$D\bar{D}^*$&9.8&\\
		&&$D^*\bar{D}^*$&23.0&&&&$D^*\bar{D}^*$&15.1&\\
		&&$D_s\bar{D}_s$&0.4&&&&$D_s\bar{D}_s$&0.0&\\
		&&$D_s\bar{D}^*_s$&0.9&&&&$D_s\bar{D}^*_s$&0.8&\\
		&&$D^*_s\bar{D}^*_s$&0.5&&&&$D^*_s\bar{D}^*_s$&0.4&\\
		&&$D\bar{D}_1$&10.0&&&&$D\bar{D}_1$&0.0&\\
		&&$D\bar{D}^{\prime}_1$&0.1&&&&$D\bar{D}^{\prime}_1$&0.0&\\
		&&$D\bar{D}^*_0$&  0.0&&&&$D\bar{D}^*_0$&4.9&\\
		&&$D\bar{D}^*_2$&4.3&&&&$D\bar{D}^*_2$&10.6&\\
		&&$D^*\bar{D}^*_0$&0.6&&&&$D^*\bar{D}^*_0$&0.0&\\
		&&&[45.5]&$[-]$&&&&[41.5]&$[-]$\\
	\bottomrule[1pt]\bottomrule[1pt]
	\end{tabular}
\end{table*}

\begin{table*}
	\caption{\label{P-wavedecay}Two-body open-charm strong decay widths of different channels for $2P$ and $3P$ charmonium states in units of MeV. $\Gamma_{\rm th}$ and $\Gamma_{\rm exp}$ denote our results and experimental measurements, respectively. The values in the square brackets mean the total widths. When the experimental datum is not available, a symbol ``$-$'' is presented.}
	\begin{tabular}{c  c  l  c  c  c  c  l  c  c }\toprule[1pt]\toprule[1pt]
		Meson &State&Channel& $\Gamma_\text{th}$&$\Gamma_\text{exp}$   &Meson &State&Channel& $\Gamma_\text{th}$&$\Gamma_\text{exp}$\\
		\hline
			$\chi_{c0}(3860)$&   $2{^3}P_0$&  $D\bar{D}$&    16.6&      &$\chi_{c1}(3872)$&   $2{^3}P_1$&   $D\bar{D}^*$&     $0.0$&        \\
			
			&    & &$[16.6]$&$[201^{+154+88}_{-\,\,\,67-82}]$ & &  &  &$[0.0]$&    $[1.19\pm0.21]$               \\
			
			$\chi_{c2}(3930)$ &     $2{^3}P_2$&   $D\bar{D}$&    $3.6$&     &$h_{c}(2P)$&     $2{^1}P_1$&    $D\bar{D}^*$&             $20.3$&       \\
			&&  $D\bar{D}^*$&  $26.9$ &     &             &               &                                           \\
			&&   &$[30.4] $  &$[35.2\pm2.2]$& &  &  &$[20.3]$&$[-]$  \\
			$\chi_{c0}(3P)$ &$3{^3}P_0$&$D\bar{D}$&12.6&&$\chi_{c1}(3P)$&$3{^3}P_1$&$D\bar{D}$&0.0&\\
			&&$D\bar{D}^*$& 0.0&&&&$D\bar{D}^*$&25.9&\\
			&&$D^*\bar{D}^*$&2.9&&&&$D^*\bar{D}^*$&3.3&\\
			&&$D_s\bar{D}_s$&0.2&&&&$D_s\bar{D}_s$& 0.0&\\
			&&$D_s\bar{D}^*_s$& 0.0&&&&$D_s\bar{D}^*_s$&1.2&\\
			&&$D^*_s\bar{D}^*_s$&0.0&&&&$D^*_s\bar{D}^*_s$&0.0&\\
			&&&[15.7]&$[-]$&&&&[30.4]&$[-]$\\
			$\chi_{c2}(3P)$ &$3{^3}P_2$&$D\bar{D}$&0.4&&$h_c(3P)$&$3^1P_1$&$D\bar{D}$&0.0&\\
			&&$D\bar{D}^*$&7.4&&&&$D\bar{D}^*$&22.8&\\
			&&$D^*\bar{D}^*$&15.2&&&&$D^*\bar{D}^*$&6.9&\\
			&&$D_s\bar{D}_s$&0.8&&&&$D_s\bar{D}_s$&0.0&\\
			&&$D_s\bar{D}^*_s$&0.0&&&&$D_s\bar{D}^*_s$&0.5&\\
			&&$D^*_s\bar{D}^*_s$&0.7&&&&$D^*_s\bar{D}^*_s$&0.7&\\
			&&&[24.4]&$[-]$&&&&[30.9]&$[-]$\\
			\bottomrule[1pt]\bottomrule[1pt]
	\end{tabular}
\end{table*}

\begin{table}
	\caption{\label{F-wavedecay}Two-body open-charm strong decay widths of different channels for $1F$ charmonium states in units of MeV. $\Gamma_{\rm th}$ and $\Gamma_{\rm exp}$ denote our results and experimental measurements, respectively. The values in the square brackets mean the total widths. When the experimental datum is not available, a symbol ``$-$'' is presented.}
	\begin{tabular}{c  c  l  c  c  c  }\toprule[1pt]\toprule[1pt]
		Meson &State&Channel& $\Gamma_\text{th}$&$\Gamma_\text{exp}$   \\\hline
		$\chi_2(1F)$ &$1{^3}F_2$&$D\bar{D}$&14.9&\\
		&&$D\bar{D}^*$&37.6\\
		&&$D^*\bar{D}^*$&0.2\\
		&&$D_s\bar{D}_s$&2.5\\
		&&&[55.1]&[$-$]\\

$\chi_3(1F)$&$1{^3}F_3$&$D\bar{D}$& 0.0\\		
&&$D\bar{D}^*$&43.6&\\
&&$D^*\bar{D}^*$&0.0&\\
&&$D_s\bar{D}_s$&0.0&\\
&&&[43.6]&[$-$]\\
		$\chi_4(1F)$ &$1{^3}F_4$&$D\bar{D}$&18.9&\\
		&&$D\bar{D}^*$&5.4&\\
		&&$D^*\bar{D}^*$&0.0&\\
		&&$D_s\bar{D}_s$&0.0&\\
		&&&[24.3]&[$-$]\\
$h_{c}(1F)$&$1{^1}F_3$&$D\bar{D}$&0.0&\\
&&$D\bar{D}^*$&33.1&\\
&&$D^*\bar{D}^*$&0.0&\\
&&$D_s\bar{D}_s$&0.0&\\
&&&[33.1]&[$-$]\\
		\bottomrule[1pt]\bottomrule[1pt]
	\end{tabular}
\end{table}

According to the heavy quark spin symmetry, the spin singlet ($S_{c\bar{c}}=0$) and spin triplet ($S_{c\bar{c}}=1$) charmonium states with the same radial number are approximately degenerate. The properties of these states may have similar features. However, the channel coupling probably has effects on the features because of their different quantum numbers.

In this section, we present numerical results about the masses, strong decay widths, and di-electric decay widths of the charmonium states. For the low-lying states, one can interpret their properties well in both quenched potential models and coupled-channel models \cite{Barnes:2005pb,Barnes:2007xu,Pennington:2007xr,Li:2009ad}. However, a significant number of charmonium states above or near relevant $S$-wave open-charm thresholds cannot be easily understood in the conventional potential model. Consideration of the coupled channel effects in the calculation becomes indispensable in understanding the spectrum of excited $c\bar{c}$ states.

We collect our numerical results for the masses of the considered charmonium states in table \ref{mass} where the highest partial wave is $L=3$ and the highest radial excitation is $3$. The first column displays charmonium states and their corresponding quantum numbers. The contributions to the mass shifts from different couple channels are shown from the second column to the seventh column. The eighth column gives the total mass shifts. The bare masses and estimated physical masses are listed in the ninth and tenth columns, respectively. For a comparison study, we also present the results in a NR potential model adopted in Ref. \cite{Barnes:2005pb} in the eleventh column. The twelfth column is for the experimental values. The last column gives the probabilities ($P_{c\bar{c}}$) of the $c\bar{c}$ components in physical states. In Fig.\ref{charmonium}, we illustrate the relative positions for the charmonia and relevant meson-meson thresholds. The calculated open-charm decay widths for $S$-, $D$-, $P$-, and $F$-wave states are listed in tables \ref{S-wavedecay}, \ref{D-wavedecay}, \ref{P-wavedecay}, and \ref{F-wavedecay}, respectively.

We will mainly concentrate on charmonium states having open-charm strong decay channels in the following subsection discussions. We will use the standard symbols to denote names of charmonium states in the quark model. If the mass appears in the meson name, it represents the observed/assigned state with corresponding $^{2S+1}L_J$ notation in this study. For the low-lying $\eta_c(1S)$, $J/\psi(1S)$, $\eta_c(2S)$, $h_c(1P)$, and $\chi_{cJ}(1P)$ ($J=0,1,2$) states, their decays into open-charm channels are all kinematically forbidden and their masses we obtain are close to the measured values. The probabilities of the $c\bar{c}$ component in these states are all above 90\%. For the $\eta_{c}(1D)$ charmonium, it has not been observed yet. Its mass is predicted to be around 3.78 GeV, a value consistent with the nonrelativestic quark model calculation of Ref. \cite{Barnes:2005pb}. Its decay into $D\bar{D}^*$ is also kinematically forbidden. For the $\psi_2(3823)$ state \cite{Belle:2013ewt}, the obtained mass is about 40 MeV smaller than the PDG value. Its strongly decay into $D\bar{D}$ is forbidden because of quantum numbers.

\subsection{The pseudoscalar charmonium states}

This type charmonium states having open-charm decay channels are $\eta_c(3S)$ and $\eta_c(4S)$. Both of them have not been observed experimentally. The predicted mass of the former state is 4017 MeV and the latter 4345 MeV, which are consistent with the results obtained in the NR potential model \cite{Barnes:2005pb}. From table \ref{mass}, the coupled-channel effects affect the $\eta_c(3S)$ state more than the $\eta_c(4S)$, which leads to the small proportion (61\%) of the $c\bar{c}$ component in the physical $\eta_c(3S)$. From table \ref{S-wavedecay}, both states seem not to be broad.

\subsection{ The vector charmonium states }

For the vector charmonium states, as mentioned above, the $S$-$D$ wave mixing effects should be considered. We do not include such effects when presenting the results in tables \ref{mass} and \ref{D-wavedecay}. After such effects are considered, the redetermined masses, di-electric decay widths, and two-body open-charm strong decay widths are summarized in tables \ref{massmix}, \ref{Di-leptonic}, and \ref{S-Dmixdecay}, respectively.

\renewcommand\tabcolsep{0.13cm}
 \begin{table}[!htbp]
\caption{\label{massmix}Masses of charmonium states in the $J^{PC}=1^{--}$ family after the $S$-$D$ wave mixing effects are considered in the adopted coupled-channel model. The first column shows the observed states while the second column lists mixed charmonia defined in Eq. \eqref{mixtureangle}. The $\delta m$ and $M_A$ have the same meanings as table \ref{mass} but the values of $\delta m$ are different. The $M_{S-D}=M_A+\delta m$ denotes the calculated masses. The experimental data $M_{exp}$ are taken from the particle data book \cite{ParticleDataGroup:2022pth}. All the masses are given in units of MeV.}
\begin{tabular*}{1.0\columnwidth}{@{\extracolsep{\fill}}cccccc@{}}\toprule[1pt]\toprule[1pt]
Meson & States                                                 &$\delta m$     &$M_A$           &$M_{S-D}$   &   $M_{exp}$\\ \hline
$\psi(3686)$&$\tilde{\psi}(2S)$       &$-60.8$  &3728.9                 &3668.1          &   $3686.10\pm0.06$\\
$\psi(3770)$&$\tilde{\psi}(1D)$       &$-76.9$  &3843.3                 &3766.3          &   $3773.7\pm0.4$\\
$\psi(4040)$&$\tilde{\psi}(3S)$       &$-78.5$ &4107.3                  &4028.8          &    $4039\pm 1$\\
$\psi(4160)$&$\tilde{\psi}(2D)$      &$-37.0$  &4169.3                   &4132.3          &$4191\pm5$\\
$\psi(4360)$&$\tilde{\psi}(4S)$      &$-72.6$   &4417.5                 &4344.9          &$4372\pm9$\\
$\psi(4415)$ &$\tilde{\psi}(3D)$        &$-32.6$   &4453.4                &4420.8          & $4421\pm4$       \\
\bottomrule[1pt]\bottomrule[1pt]
\end{tabular*}
\end{table}

\begin{table}[!htbp]
\caption{\label{Di-leptonic}Di-electric decay widths of charmonium states in the $J^{PC}=1^{--}$ family after the $S$-$D$ wave mixing effects are considered in the coupled-channel model. $S$- and $D$-wave probabilities for the $c\bar{c}$ component are given in the second and third columns, respectively. The fourth column is for the mixing angle. The calculated and measured widths \cite{ParticleDataGroup:2022pth} in units of keV are shown in the last two columns, respectively.}
\begin{tabular*}{1.0\columnwidth}{@{\extracolsep{\fill}}cccccc@{}}\toprule[1pt]\toprule[1pt]
Meson     &$P_S(\%)$  & $P_D(\%)$   &$\theta(degree) $ &$\Gamma_{ee}^{th}$&$\Gamma_{ee}^{exp}$ \\ \hline
$\psi(3686)$ &95.6  &$4.4$          &$-12.2^\circ$     &         3.03&$2.33\pm0.04$         \\
$\psi(3770)$ &$3.3$ &$96.7$            & $-10.5^\circ$  &        0.277&$0.262\pm0.018$      \\
$\psi(4040)$ &$84.2$  &$15.8$           &$-23.4^\circ$    &       1.61&$0.86\pm0.07$       \\
$\psi(4160)$ &$19.1$ &$80.9$           &$-25.9^\circ$   &        0.559&$0.48\pm0.22$        \\
$\psi(4360)$ &79.3 &20.7          &$-27.1^\circ$  &          0.899&$-$                           \\
$\psi(4415)$ &$17.3$ &$82.7$          &$-24.6^\circ$ &          0.500&$0.58\pm0.07$        \\
\bottomrule[1pt]\bottomrule[1pt]
\end{tabular*}
\end{table}

\begin{table*}
\caption{\label{S-Dmixdecay}Two-body open-charm strong decay widths of different channels for the $J^{PC}=1^{--}$ charmonium states in units of MeV: the case after $S$-$D$ mixing effects are considered. $\Gamma_{\rm th}$ and $\Gamma_{\rm exp}$ denote our results and experimental measurements, respectively. The values in the square brackets mean the total widths. When the experimental datum is not available, a symbol ``$-$'' is presented. }
\begin{tabular}{c  c  l  c  c  c  c  l  c  c }\toprule[1pt]\toprule[1pt]
Meson &State&Channel& $\Gamma_\text{th}$&$\Gamma_\text{exp}$   &Meson &State&Channel& $\Gamma_\text{th}$&$\Gamma_\text{exp}$  \\\hline
	$\psi(3686)$ &$\tilde\psi(2S)$&$D^+{D}^-$&0.0&&$\psi(3770)$&$\tilde\psi(1D)$&$D^+{D}^-$&9.1&\\
		&&$D^0\bar{D}^0$&0.0&&&&$D^0\bar{D}^0$&13.5&\\
		&&&[0.0]&$[-]$&&&&[22.5]&$[27.2\pm1.0]$\\
		$\psi(4040)$ &$\tilde\psi(3S)$&$D\bar{D}$&6.4&&$\psi(4160)$&$\tilde\psi(2D)$&$D\bar{D}$&11.8&\\
		&&$D\bar{D}^*$&9.7&&&&$D\bar{D}^*$&13.2&\\
		&&$D^*\bar{D}^*$&21.8&&&&$D^*\bar{D}^*$&20.0&\\
		&&$D_s\bar{D}_s$&0.1&&&&$D_s\bar{D}_s$&0.2&\\
		&&&&&&&$D_s\bar{D}^*_s$&1.0&\\
		&&&[38.2]&$[80\pm10]$&&&&[46.3]&$[70\pm10]$\\
		$\psi(4360)$ &$\tilde\psi(4S)$&$D\bar{D}$&0.6&&$\psi(4415)$&$\tilde\psi(3D)$&$D\bar{D}$&1.7&\\
  &&$D\bar{D}^*$&10.5&&&&$D\bar{D}^*$&8.9&\\
   &&$D^*\bar{D}^*$&12.1&&&&$D^*\bar{D}^*$&15.7&\\
  &&$D_s\bar{D}_s$&0.4&&&&$D_s\bar{D}_s$&0.8&\\
 &&$D_s\bar{D}^*_s$&0.0&&&&$D_s\bar{D}^*_s$&0.0&\\
 &&$D^*_s\bar{D}^*_s$&0.4&&&&$D^*_s\bar{D}^*_s$&0.5&\\
 &&$D\bar{D}_1$&6.9&&&&$D\bar{D}_1$&6.8&\\
 &&$D\bar{D}^{\prime}_1$&1.6&&&&$D\bar{D}^{\prime}_1$&1.2&\\
 &&$D\bar{D}^*_0$&0.0&&&&$D\bar{D}^*_0$&  0.0&\\
 &&$D\bar{D}^*_2$&5.1&&&&$D\bar{D}^*_2$&3.7&\\
 &&$D^*\bar{D}^*_0$&0.8&&&&$D^*\bar{D}^*_0$&2.3&\\
 &&&[38.6]&$[115\pm13]$&&&&[41.6]&$[62\pm20]$\\
		\bottomrule[1pt]\bottomrule[1pt]
\end{tabular}
\end{table*}

\subsubsection{$\psi(3770)$}

The $\psi(3770)$ state is the lowest charmonium above the $D \bar D$ threshold and decays dominantly into $D\bar{D}$ through $P$ wave. If it is a pure $^3D_1$ $c\bar{c}$ state, the estimated di-electric width is not consistent with experiments \cite{Kuang:1989ub,Ding:1991vu,Rosner:2001nm,Badalian:2017nyv}. To solve this problem, the $\psi(3770)$ is usually interpreted as a mixture of $\psi(2S)$ and $\psi(3D)$.

From our results in tables \ref{mass} and \ref{massmix}, the masses of $\psi(2S)$ and $\psi(3770)$ are just slightly changed after the $S$-$D$ wave mixing is considered. The extracted mixing angle from the coupled channel model is $\theta=-10.5^{\circ}$. This value is in agreement with Ref. \cite{Kuang:1989ub} where $\theta=-10^{\circ}$ is obtained. Using $\theta=-10.5^{\circ}$, we obtain $\Gamma_{ee}^{th}=0.277$ keV for the di-electric width of $\tilde{\psi}(1D)$, which is in excellent consistent with the measured value $\Gamma_{ee}^{exp}=0.262\pm0.018$ keV for $\psi(3770)$ \cite{ParticleDataGroup:2022pth}. Assuming that the $\psi(3770)$ state is the $\tilde{\psi}(1D)$ charmonium, we get $\Gamma_{\rm th}=22.5$ MeV for the open-charm strong decay width. It is close to the experimental result of $\Gamma=27.2\pm1.0$ MeV. Therefore, it is reasonable to assign the $\psi(3770)$ as the $\tilde{\psi}(1D)$ mixed state from our results of mass and widths. If we estimate the branching ratio for its non-$D\bar{D}$ decay, a value about 17\% is obtained.

 \subsubsection{$\psi(4040)$}
 
The $\psi(4040)$ state is generally assigned to the pure $3^3S_1$ charmonium in the quenched potential models \cite{Barnes:2005pb}. However, the predicted mass in the NR potential model \cite{Barnes:2005pb} is about 30 MeV higher than the experimental value. It is also necessary to discuss the effects of $S$-$D$ mixing and channel coupling on the properties of $\psi(4040)$.

Comparing tables \ref{mass} and \ref{massmix}, one sees that the mixing between $S$- and $D$-wave charmonia induces $\sim$15 MeV mass shift for the $3^3S_1$ state. The mass of $\tilde{\psi}(3S)$ (4028.8 M3V) is consistent with the measured value $4039\pm1$ MeV for the $\psi(4040)$. The extracted mixing angle for the $\tilde{\psi}(3S)$ state is found to be $\theta=-23.4^{\circ}$ which is close to $(-28_{-2}^{+1})^{\circ}$ obtained in Ref. \cite{Badalian:2017nyv}. With this mixing angle, the calculated di-electric width $\Gamma_{ee}^{th}=1.61$ keV is consistent with the Belle data $\Gamma_{ee}^{exp}=1.6\pm0.3$ keV \cite{Uglov:2016orr}. In Ref. \cite{Fu:2018yxq}, treating the $\psi(4040)$ as the $\tilde{\psi}(3S)$ state, the authors obtained a strong decay width $86.9$ MeV by using the $^3P_0$ model. However, our result for the open-charm strong decay width is $\Gamma=38.2$ MeV, a smaller number than the measured total width $80\pm10$ MeV \cite{ParticleDataGroup:2022pth}. The reason is that we adopted a smaller $\gamma$ in the $^3P_0$ model. In Ref. \cite{Deng:2023mza}, a larger decay width of $\psi(4040)$ is obtained with $\gamma=0.422$. If the $\psi(4040)$ corresponds to the $\tilde{\psi}(3S)$ state, its dominant decay modes should be $D\bar{D}$, $D\bar{D}^*$, and $D^*\bar{D}^*$. From above results, the $\psi(4040)$ may be assigned as the $\tilde{\psi}(3S)$ mixed charmonium state, although its theoretical open-charm decay width in our model is about half of the experimental width. Future measurements on the branching ratios will be helpful to test the model or to find an appropriate $\gamma$.

\subsubsection{$\psi(4160)$ and $\psi(4230)$}

The $\psi(4160)$ state was initially observed in the $e^+e^-$ annihilation process forty years ago \cite{DASP:1978dns}. Its mass and width from PDG \cite{ParticleDataGroup:2022pth} are $4191\pm5$ MeV and $70\pm10$ MeV, respectively. The mass is consistent with the quark potential model result for the $2^3D_1$ charmonium state. In 2005, the BaBar Collaboration reported the observation of $Y(4260)$ in the $e^+e^-\to \pi^+\pi^- J/\psi$ process. With the improved measurements \cite{BESIII:2016adj,BESIII:2016bnd}, the $Y(4260)$ is reconstructed as two states $\psi(4230)$ and $\psi(4360)$. Then, the $Y(4260)$ was renamed to be $\psi(4230)$ in the particle data book with $M=4222.7\pm2.6$ MeV and $\Gamma=49\pm8$ MeV. Very recently, the BESIII Collaboration precisely measured the Born cross sections of $e^+e^-\to D^{*+}_s D^{*-}_s$ \cite{BESIII:2023wsc} and observed two resonances. One of them is consistent with $\psi(4415)$. The mass and width of the other resonance from the fit are $M=4186.8\pm8.7\pm30$ MeV and $\Gamma=55\pm15\pm53$ MeV, respectively. If one considers the systematic uncertainties, this state is compatible with both $\psi(4230)$ and $\psi(4160)$. Using a coupled-channel scheme with unitarity, the authors of Ref. \cite{Zhou:2023yjv} extracted the pole position ($\sqrt{s}=4222-32i$ MeV) of the $\psi(4160)$ by analyzing the cross section data of $e^+e^-\to D^{(*)}\bar{D}^{(*)}$ and $D\bar{D}\pi$ processes. Their results imply that the $\psi(4160)$ and $\psi(4230)$ might be the same state, $c\bar{c}(2^3D_1)$ mixed with continuum states. In addition, in Ref. \cite{Nakamura:2023obk}, the authors studied the $e^+e^-\to c\bar{c}$ processes based on a global coupled-channel analysis. They found that two pole positions are around 4230 MeV, but did not find a $\psi(4160)$ pole.
 
In our scheme, similar to the $\psi(3770)$ case, the $\psi(4160)$ is assumed as the $3S$-$2D$ mixing state. If one treats it as  the pure $2^3D_1$ state, the predicated di-electric width is about 14 times smaller than the experimental measurement \cite{Badalian:2007ore}. The mixing between $\psi(3S)$ and $\psi(2D)$ can be used to understand its measured di-electric width.

From table \ref{massmix}, the mass of $\tilde{\psi}(2D)$ in our calculation is $M=4132.2$ MeV which is in accordance with the NR model result $M=4142$ MeV \cite{Barnes:2005pb} but is about 60 MeV smaller than the PDG mass of $\psi(4160)$. The extracted mixing angle between the $3^3S_1$ and $2^3D_1$ states is $-25.9^{\circ}$. Using this angle, one obtains a di-electric width $\Gamma_{ee}^{th}=0.56$ keV which is in agreement with the PDG value $\Gamma_{ee}^{exp}=0.48\pm0.22$ keV for the $\psi(4160)$ \cite{ParticleDataGroup:2022pth}. From the open-charm decay width $46.3$ MeV which is consistent with the BESIII result $55\pm15\pm53$ MeV \cite{BESIII:2023wsc}, assigning the $\psi(4160)$ as the $\tilde{\psi}(2D)$ mixed charmonium is reasonable. The consistency between our width and the experimental measurements indicates that the open-charm channels dominate its decay if the $\psi(4160)$ is assigned as the $\tilde{\psi}(2D)$ state. To understand its structure, ratios between partial widths of different channels play an important role. From table \ref{S-Dmixdecay}, the ratio between $D\bar{D}$ and $D^*\bar{D}^*$ for the $\tilde{\psi}(2D)$ charmonium is around $\Gamma(D\bar{D}):\Gamma(D^*\bar{D}^*)=0.59$. This value is somewhat larger than the BaBar result $0.02\pm0.04$ \cite{BaBar:2009elc}. The ratio $\Gamma(D\bar{D}^{*}):\Gamma(D^*\bar{D}^*)=0.66$ is close to the upper limit of  the measured value $0.34\pm0.15$ \cite{BaBar:2009elc}. Therefore, the above comparison indicates that the $\psi(4160)$ is a reasonable candidate of the $\tilde{\psi}(2D)$ state, although there are still inconsistencies at present. In our case, the $\psi(4230)$ cannot be interpreted as a $c\bar{c}$ state. Further investigations on the natures of $\psi(4160)$ and $\psi(4230)$ are certainly needed.

\subsubsection{$\psi(4360)$}

The $\psi(4360)$ state was first observed in the initial-state radiation (ISR) process $e^+e^-\rightarrow\gamma_{ISR}\pi^+\pi^-\psi(2S)$ by the BaBar Collaboration in 2006 \cite{BaBar:2006ait}. It was subsequently confirmed by the Belle Collaboration \cite{Belle:2007umv}. Because of its production process, the quantum numbers of $\psi(4360)$ are naturally $J^{PC}=1^{--}$. The $\psi(4360)$ was interpreted as the canonical $\psi(3D)$ charmonium state in the NR potential model \cite{Barnes:2005pb}, while the estimated mass is about 80 MeV larger than the experimental mass \cite{Ding:2007rg}. It is also suggested that the $\psi(4360)$ may be a candidate of $\psi(4S)$ or $\psi(3D)$ in the screened confinement potential model \cite{Segovia:2008zz,Li:2009zu}. Moreover, the $\psi(4360)$ can be assigned as a molecular state \cite{Chen:2017abq}, tetraquark state \cite{Anwar:2018sol,Wang:2018rfw}, or a hybrid state \cite{Miyamoto:2018zfr}.

Similarly to other $1^{--}$ states, we check whether the $\psi(4360)$ can be assigned as the $\tilde{\psi}(4S)$ mixed charmonium state based on our results. The assignment for the $\psi(3D)$ charmonium will be discussed soon. In our coupled-channel model, the predicted mass ($4344.9$ MeV) of $\tilde{\psi}(4S)$ is consistent with the $\psi(4360)$ ($4372\pm9$ MeV) \cite{ParticleDataGroup:2022pth}. For a higher charmonium, large $S$-$D$ mixing angle can usually be used to understand its measured di-electric width \cite{Badalian:2008dv,Anwar:2016mxo,Badalian:2017nyv}. From table \ref{Di-leptonic}, the extracted mixing angle $-27.1^{\circ}$ results in the estimated di-electric decay width of $\tilde{\psi}(4S)$ to be $\Gamma_{ee}^{th}=0.90$ keV. This value as a criterion for the assignment of $\psi(4360)$ can be tested by future experiments. It is worth noting that our result is in agreement with $\Gamma_{ee}^{th}=0.78$ keV obtained in Ref. \cite{Badalian:2008dv}. For the open-charm decay widths, our result (38.6 MeV) of $\tilde{\psi}(4S)$ is much smaller than the measured width ($115\pm13$ MeV) of $\psi(4360)$. It has been argued in Ref. \cite{Fu:2018yxq} that the small width may be caused by the oscillation behavior of the decay amplitude and the $S$-$D$ mixing effect. In addition, the experiments have not observed the open-charm decay channels of $\psi(4360)$ at present. Assigning this state as the $\tilde{\psi}(4S)$ mixed charmonium is possible. Whether the small open-charm decay width of $\psi(4360)$ is reasonable or not needs to be answered by future measurements. The width ratios between different open-charm channels determined with our results in table \ref{S-Dmixdecay} may also be used to test the assignment of $\psi(4360)$ as the $\tilde{\psi}(4S)$ charmonium.

 \subsubsection{$\psi(4415)$}

The $\psi(4415)$ is usually interpreted as the $\psi(4S)$ charmonium state. However, the di-electronic width obtained in the potential model with this assignment is about two times larger than the experimental value \cite{Badalian:2008dv}. Here, we check the possibility to assign the $\psi(4415)$ to be the $\tilde{\psi}(3D)$ charmonium by taking into account the $S$-$D$ mixing contribution. This assignment is helpful for us to understand the di-electric decay width of the $\psi(4415)$.

As shown in table \ref{massmix}, the predicted mass $M(\tilde{\psi}(3D))=4420.8$ MeV in our coupled-channel model is in good agreement with the measured mass $4421\pm4$ MeV of $\psi(4415)$ \cite{ParticleDataGroup:2022pth}. With the extracted mixing angle $-24.6^{\circ}$, one obtains the di-electric width $\Gamma_{ee}^{th}(\tilde{\psi}(3D))=0.5$ keV. It is also consistent with the experimental result $\Gamma_{ee}^{exp}(\psi(4415))=0.58\pm0.07$ keV \cite{ParticleDataGroup:2022pth}. The open-charm decay width of $\tilde{\psi}(3D)$ is found to be 41.6 MeV in the adopted coupled-channel model. It is consistent with the measured total width of $\psi(4415)$  ($62\pm20$ MeV \cite{ParticleDataGroup:2022pth}), but much smaller than the recent BESIII result ($122.5\pm7.5\pm8.1$ MeV \cite{BESIII:2023wsc}). To understand the nature of $\psi(4415)$, ratios of partial widths in different assignment schemes would play an import role. The $\tilde{\psi}(3D)$ dominantly decays into the $D\bar{D}^*$, $D^*\bar{D}^*$, $D\bar{D}_1$, and $D\bar{D}^*_2$ channels. The partial width ratio between $D\bar{D}$ and $D^*\bar{D}^*$ is estimated to be $\Gamma(D\bar{D}):\Gamma(D^*\bar{D}^*)=0.11$ from our results. This number agrees with BaBar $(0.14\pm0.12\pm0.03)$ \cite{BaBar:2009elc}. The ratio between the $D\bar{D}^*$ and $D^*\bar{D}^*$ channels, 0.56, is somewhat larger than the experimental value $0.17\pm 0.25\pm0.03$ \cite{BaBar:2009elc}. The predicted branching fraction $Br(\psi(4415)\to D\bar{D}^*_2)\sim9\%$ is also compatible with the Belle data $(\sim10\%)$ \cite{Belle:2007xvy}. From the above analysis, the $\psi(4415)$ is a good candidate of the $\tilde{\psi}(3D)$ mixed charmonium state, which confirms the conclusion drawn in Ref. \cite{Segovia:2008zz}. One should note that the dominant component in $\psi(4415)$ in this scheme is the $D$-wave $c\bar{c}$ rather than the $S$-wave $c\bar{c}$. If $\psi(4415)$ is interpreted as the $S$-wave dominant $c\bar{c}$ state, there is no room for the $\psi(4360)$ from our model calculation.

\subsection{The spin-2 and -3 $D$-wave charmonium states}

Besides the above $J^P=1^{--}$ $D$-wave dominant charmonia $\psi(3770)$, $\psi(4160)$, and $\psi(4415)$, we also studied seven higher spin $D$-wave states. One understands the nature of the observed $\psi_3(3842)$ with the $\psi_3(1D)$ charmonium. The related exotic state of $\eta_c(2D)$ is the $X(4160)$. The properties of other five states are our predictions.

\subsubsection{$\psi_3(3842)$}

Recently, the LHCb reported the observation of a new narrow charmonium state $\psi_{3}(3842)$ in the decay channels $\psi_3(3842)\to D^0\bar{D}^0$ and $\psi_3(3842)\to D^+D^-$ with high statistical significance \cite{LHCb:2019lnr}. This state has a mass $M=3842.71\pm0.16\pm0.12$ MeV and a width $2.79\pm0.51\pm0.35$ MeV. It is assigned as the $\psi_3(1D)$ charmonium state with $J^{PC}=3^{--}$. We check this assignment based on our results.

In the coupled-channel model, we get a mass $M=3805.6$ MeV for the $\psi_3(1D)$ charmonium. It is consistent with the result of the NR potential model (3806 MeV) \cite{Barnes:2005pb}. From table \ref{D-wavedecay}, the open-charm strong decay width (1.9 MeV) of $\psi_3(1D)$ is close to the lower limit of the measured width of $\psi_3(3842)$. With the present information, one may consider the $\psi_3(3842)$ as the $\psi_3(1D)$ charmonium, although the obtained mass of $\psi_{3}(1D)$ is about 37 MeV smaller than the $\psi_3(3842)$. This assignment also indicates that the $D\bar{D}$ mode should dominate its decay. Future experimental measurements on decays would provide useful information. The partial width ratio between the two dominant decay channels $D^+D^-$ and $D^0\bar{D}^0$ we get is 0.73. This ratio is in accordance with a recent theoretical result 0.84 \cite{Wang:2022dfd} by using the BS method and the $^3P_0$ model. Based on its mass and decay properties, the $\psi_3(3842)$ is a good candidate of the $\psi_3(1D)$ charmonium state.

\subsubsection{$X(4160)$}

The $X(4160)$ was first observed in the $D^{(*)}D^{(*)}$ invariant mass spectrum of the $e^+e^-\to J/\psi D^{(*)}D^{(*)}$ processes by the Belle \cite{Belle:2007woe}. In 2021, the LHCb announced a $J/\psi\phi$ resonance consistent with the $X(4160)$ \cite{LHCb:2021uow}. The preferred quantum numbers for this resonance are $J^{PC}=2^{-+}$. If they are really the same state, its mass and width are $4153^{+23}_{-21}$ MeV and $136^{+60}_{-35}$ MeV, respectively \cite{ParticleDataGroup:2022pth}.

From table \ref{mass}, the calculated mass of $\eta_{c}(2D)$ is 4136.2 MeV which is consistent with Ref. \cite{Cao:2012du} and the measured $X(4160)$. From table \ref{D-wavedecay}, the corresponding two-body open-charm decay width is 42.5 MeV which is smaller than the width of $X(4160)$. If the dominant decay mode of such a state is not $D^{(*)}\bar{D}^{(*)}$, the possible assignment of $X(4160)$ as the $\eta_{c}(2D)$ cannot be excluded. At present, the Belle result $\Gamma(D\bar{D}):\Gamma(D\bar{D}^*)<0.09$ does not conflict with the fact that the decay channel $D\bar{D}$ is forbidden for the $\eta_{c}(2D)$ state. However, the measured partial width ratio $\Gamma(D\bar{D}^*):\Gamma(D^*\bar{D}^*)=0.22$ \cite{ParticleDataGroup:2022pth} is much smaller than the expected $\Gamma(D\bar{D}^*):\Gamma(D^*\bar{D}^*)=2$. Moreover, the inconsistency for the production rates of the $\eta_c(2D)$ in $e^+e^-$ annihilation between experimental data and theoretical expectations does not favor this assignment \cite{Chao:2007it}. We need more information to understand the nature of $X(4160)$.

\subsubsection{Other $2D$ and $3D$ states}

In the remaining five states, from table \ref{mass}, two are $2D$ charmonia $\psi_2(2D)$ and $\psi_3(2D)$. They are all around 4.14 GeV which are in accordance with the predictions with the NR potential model \cite{Barnes:2005pb}. This degeneration feature is easy to understand since they are spin partner states of the $\psi(4160)$ in the heavy quark limit \cite{Casalbuoni:1992yd}. The predicted width of the $\psi_2(2D)$ state is about $55$ MeV and it dominantly decays into $D\bar{D}^*$ and $D^*\bar{D}^*$. The ratio between partial widths is $\Gamma(D\bar{D}^*):\Gamma(D^*\bar{D}^*)=1.9:1.0$. The $\psi_3(2D)$ state has a narrower width 20.3 MeV and its main decay modes are $D\bar{D}$, $D\bar{D}^*$, and $D^*\bar{D}^*$. We predict its ratios between partial widths to be $\Gamma(D\bar{D}):\Gamma(D\bar{D}^*):\Gamma(D^*\bar{D}^*)=1.0:2.0:3.9$.

The three $3D$ states $\eta_{c}(3D)$, $\psi_2(3D)$, and $\psi_3(3D)$ are spin partners of the $\psi(4415)$. Their predicted masses are all around 4.4 GeV. Their two-body open-charm strong decay widths have been shown in table \ref{D-wavedecay} where one finds that the total widths are all around 40 MeV. The total width of $\eta_{c}(3D)$ gets contributions mainly from $D\bar{D}^*$, $D^*\bar{D}^*$, $D\bar{D}^*_0$, and $D\bar{D}^*_2$ modes. The partial width ratios between these channels are predicted to be $\Gamma(D\bar{D}^*):\Gamma(D^*\bar{D}^*):\Gamma(D\bar{D}^*_0):\Gamma(D\bar{D}^*_2)=2.0:3.1:1:2.1$. The $\psi_2(3D)$ state mainly decays into $D\bar{D}^*$, $D^*\bar{D}^*$, and $D\bar{D}_2^*$. The ratios between partial widths of these channels are $\Gamma(D\bar{D}^*):\Gamma(D^*\bar{D}^*):\Gamma(D\bar{D}_2^*)=1.6:1.4:1$. The $\psi_3(3D)$ state has four dominant decay channels $D\bar{D}$, $D^*\bar{D}^*$, $D\bar{D}_1$, and $D\bar{D}^*_2$. We predict the relevant ratios to be $\Gamma(D\bar{D}):\Gamma(D^*\bar{D}^*):\Gamma(D\bar{D}_1):\Gamma(D\bar{D}^*_2)=1.3:5.4:2.3:1.0$.

\subsection{The P-wave charmonium states}
 
The involved states in the present study are $2P$ and $3P$ charmonia. From table \ref{mass}, the coupled-channel effects have relatively significant impacts on the $2P$ charmonia, especially $\chi_{c1}(2P)$. The proportion of the $c\bar{c}$ component in a physical state can even be lowered to $<50\%$. Up to now, three related charmonia ($\chi_{c0}(3860)$, $\chi_{c1}(3872)$, and $\chi_{c2}(3930)$) of the four $2P$ partner states have been observed. We expect that the $h_c(2P)$ charmonium around 3.9 GeV with a width tens of MeV will be observed in the near future. Its two-body open-charm decay mode is just $D\bar{D}^*$. Up to now, no $3P$ charmonium has been observed. The obtained mass and decay properties of them are our predictions in the employed coupled-channel model.
 
\subsubsection{$\chi_{c1}(3872)$}
 
This mysterious charmonium-like state was first observed in the decay $B\rightarrow K\pi^{+}\pi^{-}J/\psi$ by the Belle Collaboration \cite{Belle:2003nnu} and then confirmed by BaBar \cite{BaBar:2004oro}. The PDG mass and width of $\chi_{c1}(3872)$ \cite{ParticleDataGroup:2022pth} are $M=3871.65\pm0.06$ MeV and $\Gamma=1.19\pm0.21$ MeV, respectively. The $\chi_{c1}(3872)$ cannot be well interpreted by the quenched potential models \cite{Godfrey:1985xj,Barnes:2005pb} since the calculated mass of $\chi_{c1}(2P)$ is about tens of MeV larger than the experimental value. It is mainly interpreted as a $D\bar{D}^{*}$ molecule \cite{Swanson:2003tb,Braaten:2005ai,Liu:2008fh,Thomas:2008ja,Dong:2009yp,Baru:2011rs,Li:2012cs,Wang:2017dcq,Guo:2017jvc,Yang:2023mov}, a compact tetraquark state \cite{Maiani:2004vq,Hogaasen:2005jv,Chen:2016qju,Shi:2021jyr}, or a charmonium with significant continuum components \cite{Suzuki:2005ha,Meng:2007cx,Li:2009ad,Ortega:2009hj,Takizawa:2010rxa,Danilkin:2010cc,Ferretti:2013faa,Duan:2020tsx,Coito:2012vf,Tan:2019qwe}. In Ref.\cite{Kalashnikova:2009gt}, the authors calculated the spectral density using the Flatt$\acute{e}$-like parametrization. Their results showed that there is a large $D\bar{D}^{*}$ component in the $\chi_{c1}(3872)$ wave function. A number of authors got a similar conclusion. The probability of meson-meson continuum component in the $\chi_{c1}(3872)$ may be $85\%$ \cite{Takizawa:2010rxa}, $93\%$ \cite{Ortega:2009hj}, $73.0\%$ \cite{Zhou:2017dwj}, $85.3\%$ \cite{Ferretti:2018tco}, or 70\% \cite{Tan:2019qwe}, depending on the adopted model.

Now we concentrate on the $\chi_{c1}(2P)$ charmonium in our coupled-channel calculation. The bare mass (3963 MeV) is much higher than the measured value. After the coupled-channel effects are considered, we get a mass 3870.1 MeV for the $\chi_{c1}(2P)$ state which is consistent with the PDG result for the $\chi_{c1}(3872)$ \cite{ParticleDataGroup:2022pth}. It is worth noting that the mass shift provided by the $D\bar{D}^*$ channel (-57.5 MeV) is lager than contributions from other channels. In this calculation, we find that the continuum state component is around $57\%$. Because our calculated mass and the measured mass are both lower than the $D^0\bar{D}^{0*}$ threshold, the two-body open-charm strong decay width cannot be estimated with Eq. \eqref{two body decay}. Therefore, we consider its quasi two-body decay \cite{Capstick:1993kb,Roberts:1997kq,Segovia:2009zz,Gui:2018rvv,Ferretti:2014xqa,Li:2023wug}.
\begin{eqnarray}\label{three body decay}
\Gamma&=&\int_0^{q_{max}}dq\frac{2q^2\sum_{J,L}|\mathcal{M}^{JL}(\chi_{c1}(2P)\to D^{0}\bar{D}^{0*})|^2}{[M_{\chi_{c1}(2P)}-E_{\bar{D}^{0*}}(q)-E_{D^0}(q)]^2+\frac14\Gamma^2_{\bar{D}^{0*}}}\notag\\
&&\times\Gamma_{\bar{D}^{0*}\to\bar{D}^0\pi^0}.
\end{eqnarray}
Here, $\mathcal{M}^{JL}(\chi_{c1}(2P)\to D^{0}\bar{D}^{0*})$ is the decay amplitude obtained with Eq. \eqref{integral}. The  maximum relative momentum $q_{max}$ for the three-body decay $\chi_{c1}(2P)\to(\bar{D}^0\pi^0)_{\bar{D}^{*0}}D^0$ is given by
\begin{eqnarray}
q_{max}&=&\frac{\sqrt{M^2_{\chi_{c1}(2P)}-(M_{\bar{D}^0}+M_{\pi^0}+M_{D^0})^2}}{2M_{\chi_{c1}(2P)}} \notag\\
&&\times\sqrt{M^2_{\chi_{c1}(2P)}-(M_{\bar{D}^0}+M_{\pi^0}-M_{D^0})^2}.
\end{eqnarray}
We use the upper limit for the measured width of $\bar{D}^{*0}$, $\Gamma_{\bar{D}^{0*}}=2.1$ MeV \cite{ParticleDataGroup:2022pth}, to estimate the width of $\chi_{c1}(2P)$. Substituting the partial width $\Gamma_{\bar{D}^{0*}\to\bar{D}^0\pi^0}=0.65\Gamma_{\bar{D}^{0*}}$ into Eq. \eqref{three body decay}, one gets $\Gamma(\chi_{c1}(2P))=0.94$ MeV. It is in agreement with the experimental result $\Gamma(\chi_{c1}(3872))=1.19\pm0.21$ MeV. Therefore, assigning the $\chi_{c1}(3872)$ as a structure of $c\bar{c}(2^3P_1)$ core plus open-charm meson continuum is reasonable based on the obtained mass and decay width.

\subsubsection{$\chi_{c0}(3860)$ and $\chi_{c0}(3915)$}

The $\chi_{c0}(3860)$ state was observed in the $e^{+}e^{-}\rightarrow J/\psi D\bar{D}$ process by the Belle Collaboration \cite{Belle:2017egg}. Its mass and decay width are $3862^{+26+40}_{-32-13}$ MeV and $201^{+154+88}_{-\,\,\,67-82}$ MeV, respectively. In theoretical discussions, it has been assigned as the $2^3P_0$ charmonium \cite{Wang:2022dfd}, a scalar tetraquark \cite{Chen:2017dpy,Wang:2017lbl,Wu:2018xdi}, or a mixed state of the $c\bar{c}$ and continuum state \cite{Zhou:2017dwj}. Prior to the Belle's result, Ref. \cite{Guo:2012tv} had obtained a broad $\chi_{c0}(2P)$ around 3840 MeV using the Belle and BaBar's data sets.

For the $\chi_{c0}(2P)$ state in our model calculation, the mass we get (3859 MeV) is close to that of $\chi_{c0}(3860)$. This result is consistent with the NR potential model prediction of Ref. \cite{Barnes:2005pb} (3852 MeV) and the Friedrichs-model-like scheme calculation of Ref. \cite{Zhou:2017dwj} (3861 MeV). If the assignment for the $\chi_{c0}(3860)$ as $\chi_{c0}(2P)$ is correct, the state has only one two-body open-charm decay channel $D\bar{D}$ and the width is predicted to be $\Gamma=16.6$ MeV. Our result of narrow width agrees with Refs. \cite{Gui:2018rvv, Wang:2022dfd} where the adopted decay model is also the $^3P_0$ quark pair creation model. In Ref. \cite{Yu:2017bsj}, a larger decay width (110$\sim$180 MeV) for the $\chi_{c0}(2P)$ charmonium was obtained with a larger $\gamma$ within the same model. Since the available experimental decay width of the $X(3860)$ state has large uncertainty, the possibility that the $\chi_{c0}(3860)$ is the $\chi_{c0}(2P)$ charmonium is not ruled out.

The $\chi_{c0}(3915)$ was firstly observed in the $J/\psi\omega$ invariant mass distribution in the decay $B\to K\omega J/\psi$ by the Belle Collaboration \cite{Belle:2004lle}. Later, the BaBar Collaboration confirmed this state in the same process \cite{BaBar:2007vxr}. The quantum numbers of $\chi_{c0}(3915)$ are $J^{PC}=0^{++}$ and it has a mass $3921.7\pm1.8$ MeV and a width $\Gamma=18.8\pm3.5$ MeV \cite{ParticleDataGroup:2022pth}. This state can also be assigned as the $\chi_{c0}(2P)$ charmonium in some unquenched quark models \cite{Duan:2020tsx,Deng:2023mza}. In the NR potential model of Ref. \cite{Barnes:2005pb} and the screened potential model of Ref. \cite{Li:2009zu}, however, the estimated mass of the $2^3P_0$ state is around $3.85$ GeV and it is not appropriate to assign the $\chi_{c0}(3915)$ as this charmonium. The inconsistency between experimental measurements and theoretical expectations also does not support this assignment \cite{Olsen:2014maa}. In the present work, the mass of the $\chi_{c0}(2P)$ state (3859 MeV) agrees with Refs. \cite{Barnes:2005pb,Li:2009zu,Wang:2022dfd} but is about 60 MeV lower than the measured mass of $\chi_{c0}(3915)$. Just from the mass, the possibility to relate the $\chi_{c0}(2P)$ charmonium to the $\chi_{c0}(3860)$ is higher than that to the $\chi_{c0}(3915)$. If one concentrates only on the width, the possibility is reversed. Therefore, further studies to understand these two observed exotic states are still needed.

\subsubsection{$\chi_{c2}(3930)$}

The $\chi_{c2}(3930)$ was discovered in the $\gamma\gamma\rightarrow D\bar D$ process by Belle \cite{Belle:2005rte} and BaBar \cite{BaBar:2010jfn}. Its mass is $3922.5\pm1.0$ MeV and its width is $35.2\pm2.2$ MeV \cite{ParticleDataGroup:2022pth}. The quantum numbers $J^{PC}=2^{++}$ are favored for this state. The $\chi_{c2}(3930)$ is usually assigned as the $\chi_{c2}(2P)$ charmonium in the quenched models \cite{Gui:2018rvv}. From our results shown in table \ref{mass}, the mass of $\chi_{c2}(2P)$ ($3958.7$ MeV) is close to the measured mass of $\chi_{c2}(3930)$. From table \ref{P-wavedecay}, the two-body open-charm strong decay width of $\chi_{c2}(2P)$ ($30.4$ MeV) is in accordance with the width of $\chi_{c2}(3930)$. Therefore, one may assign the state as the $\chi_{c2}(2P)$ charmonium. If the $\chi_{c2}(3930)$ indeed corresponds to the $\chi_{c2}(2P)$, its decays into $D\bar{D}$ and $D\bar{D}^*$ channels have the most important contributions. The ratio between relevant partial widths we predict is  $\Gamma(D\bar{D}):\Gamma(D\bar{D}^*)=0.13$. This ratio is different from that in Ref. \cite{Wang:2022dfd}. One may use such a ratio to test different models once experimental data are available in the future.

\subsubsection{The $3P$ states}

For the four $3P$ states, their masses are all around 4.2 GeV. Based on our calculation, the mass of $\chi_{c0}(3P)$ (4205 MeV) is consistent with the mass 4202 MeV obtained in the NR quark model of Ref. \cite{Barnes:2005pb}. If its decay is saturated with the two-body open-charm channels, this charmonium mainly decays into $D\bar{D}$ and $D^*\bar{D}^*$ and the total width is about 16 MeV. Although its decay into $D_s\bar{D}_s$ is not forbidden, this channel is suppressed. The predicted ratio between the main partial widths in our model is $\Gamma(D\bar{D}):\Gamma(D^*\bar{D}^*)=4.3$. For the $\chi_{c1}(3P)$ charmonium, its mass we get is 4213 MeV which is smaller than the result of Ref. \cite{Barnes:2005pb} (4271 MeV). The total width of this state is about $30$ MeV and it dominantly decays into $D\bar{D}^*$. The channels $D^*\bar{D^*}$ and $D_s\bar{D}_s^*$ are relatively suppressed. The ratios between partial widths of these three channels are $\Gamma(D\bar{D}^*):\Gamma(D^*\bar{D^*}):\Gamma(D_s\bar{D}_s^*)=22:3:1$. Our $\chi_{c2}(3P)$ charmonium is located at the position $4276$ MeV. This number is about 41 MeV smaller than the calculated mass of Ref. \cite{Barnes:2005pb}. The width of $\chi_{c2}(3P)$ is around 25 MeV and it has two dominant decay modes $D\bar{D}^*$ and $D^*\bar{D^*}$. The width is compatible with the result of Ref. \cite{Wang:2022dfd}. The partial width ratio we get is $\Gamma(D\bar{D}^*):\Gamma(D^*\bar{D^*})=0.5$. For our singlet charmonium $h_c(3P)$, it is about 37 MeV lower than the same state in Ref. \cite{Barnes:2005pb}. The width of this state is about 31 MeV and it is consistent with the result of Ref. \cite{Wang:2022dfd} (31 MeV). The $D\bar{D}^*$ and $D^*\bar{D^*}$ decay channels are the dominant ones, with the branching fraction reaching up to $96\%$ of the two-body open-charm decay modes. The corresponding ratio between partial widths is $\Gamma(D\bar{D}):\Gamma(D^*\bar{D}^*)=3.3$. We hope that our results for the $3P$ charmonia are useful in identifying higher $P$-wave $c\bar{c}$ states in future experiments.

\subsection{The $F$-wave charmonium states}

From numerical results in our coupled-channel model, the four $1F$ charmonium states are located around 4.0 GeV. This is different from the case in the NR potential model \cite{Barnes:2005pb}. The states are around 4.23 GeV there. The mass of $\chi_{c2}(1F)$ we obtain is 4026 MeV. If the total decay width may be represented by the two-body open-charm strong decay width, it is estimated to be $55.1$ MeV. This charmonium mainly decays into $D\bar{D}$ and $D\bar{D}^*$ and the ratio between partial widths is around $\Gamma(D\bar{D}):\Gamma(D\bar{D}^*)=1.0:2.5$. The mass and width of the $\chi_{c3}(1F)$ charmonium are $3978$ MeV and $43.6$ MeV, respectively. This state only decays into $D\bar{D}^*$. The $\chi_{c4}(1F)$ charmonium has a mass around 3997 MeV. Its dominant decay channels are $D\bar{D}$ and $D\bar{D}^*$ with a corresponding partial width ratio $\Gamma(D\bar{D}):\Gamma(D\bar{D}^*)=3.5:1.0$. For the spin singlet $h_c(1F)$ charmonium, the calculated mass is 3976 MeV and our predicted width is about $33$ MeV. Its open-charm strong decay mode is mainly $D\bar{D}^*$.

\section{Summary and discussions} \label{sec4}

In recent years, quite a number of charmonium(-like) states have been observed. It is very likely that more similar states would be further observed in future. Because vector charmonia are easy to produce at the clean $e^+e^-$ colliders, experimentalists have found redundant $J^{PC}=1^{--}$ structures (e.g. ${\cal R}(3760)$, ${\cal R}(3810)$, and Y(4544) \cite{BESIII:2023fgz,BESIII:2024jzg}) that are difficult or unable to understand in the conventional quark model. If mesons with other quantum numbers are easy to produce at supposed clean colliders, a similar situation would exist. It is necessary for us to discuss the nature of the observed states in various scenarios including multiquark configuration, rescattering mechanism, coupled channel effects, and so on.

In this paper, we studied the mass spectra, two-body open-charm strong decays, and di-electric decays for the $c\bar{c}$ states within a coupled-channel model. The channel coupling effects of $D^{(*)}\bar{D}^{(*)}/D_s^{(*)}\bar{D}_s^{(*)}$ on charmonium states with the $^3P_0$ model and the $S$-$D$ wave mixing effects on vector $c\bar{c}$ are considered. We provide explicit results regarding the mass spectra, the induced mass shifts, the probability of $c\bar{c}$ components, the $S$-$D$ mixing angles, the di-electric decay widths, and the two-body open-charm decay widths for the charmonium states up to the scale around the $D_s^*\bar{D}_s^*$ threshold. From the numerical analyses, we found that:
\begin{itemize}
\item In the vector case, the $\psi(3770)$, $\psi(4040)$, $\psi(4160)$, $\psi(4360)$, and $\psi(4415)$ can be assigned as the $\tilde{\psi}(1D)$, $\tilde{\psi}(3S)$, $\tilde{\psi}(2D)$, $\tilde{\psi}(4S)$, and $\tilde{\psi}(3D)$ mixed charmonium states, respectively.
 The consideration of $S$-$D$ mixing effects on vector charmonia can interpret their experimental di-electric widths. One should note that we extracted the mixing angles from the coupled-channel model rather than from fitting the $e^+e^-$ widths. The $\psi(4040)$, $\psi(4160)$, $\psi(4360)$, and $\psi(4415)$ states have larger mixing angles than the $\psi(3770)$.

\item  The observed resonance $\psi_3(3842)$ is a good candidate of the $\psi_3(1D)$ charmonium state according to its mass and decay properties. Its unobserved $1D$ partner $\eta_{c}(1D)$ without open-charm decay also has a mass around 3.8 GeV. As a potential candidate of $\eta_c(2D)$, the $X(4160)$ is a possible partner of the $\psi(4160)$, although there are still inconsistencies between experimental data and theoretical expectations. 
 The other two not-yet-observed $2D$ states are located at 4.14 GeV with width around 20 MeV or 50 MeV. For the three $3D$ partner states of the $\psi(4415)$, their location is probably also at around 4.4 GeV and their widths are $\Gamma\sim40$ MeV.

\item Three of the four $2P$ charmonia are related to the observed states. The exotic $\chi_{c1}(3872)$ can be assigned as the $\chi_{c1}(2P)$ state with a significant continuum contribution ($\sim$57\%). If we consider the large uncertainty in the observed decay width, the possibility to assign the $\chi_{c0}(3860)$ as the $\chi_{c0}(2P)$ charmonium cannot be ruled out. However, the $\chi_{c0}(3915)$ cannot be interpreted as the $\chi_{c0}(2P)$ state in our scheme. The $\chi_{c2}(3930)$ can be well described by the $\chi_{c2}(2P)$ charmonium state. The not yet observed $2P$ charmonium $h_c(2P)$ is probably at around 3.9 GeV with a width $\sim20$ MeV. The masses of the four $3P$ charmonium states are predicted to be $M\sim4.2$ GeV and the widths are $16\sim 30$ MeV. 

\end{itemize}

For high-lying charmonium states, there are more open-charm strong decay channels than lower states. The experimentalists could search for them not only in $D^{(*)}\bar{D}^{(*)}$ channels but also in $D_s^{(*)}\bar{D}_s^{(*)}$ channels. The predicted masses, strong decay widths, possible open-charm channels, and partial width ratios in our coupled-channel model can hopefully provide useful information for future experiments in exploring new hadron states.

\begin{table*}[htbp]
	\caption{\label{diffbeta}Masses of $2P$ charmonium states obtained in the adopted coupled-channel model with fixed $\gamma$ and different $\beta$'s. See table \ref{mass} for other explanations.}
	\resizebox{\textwidth}{!}{
		\begin{tabular*}{1.8\columnwidth}{@{\extracolsep{\fill}}c c c c c c c c c c c c c@{}}
			\toprule[1pt]\toprule[1pt]\multicolumn{12}{c}{$\gamma=0.2$, $\beta=0.25$ GeV} \\\hline
			Meson & $D\bar{D}$ & $D\bar{D}^{*}$ & $D^*\bar{D}^*$ &$D_s\bar{D}_s$  &$D_s\bar{D}^*_s$ &$D^*_s{D}^*_s$&$\delta m$&$M_A$&$M_\text{cou}$&$M_\text{non}$&$M_\text{exp}$&$P_{c\bar{c}}(\%)$  \\\hline
			$h_{c}(2P)(2{^1}P_1)$ &$0.0$&$-51.2$&$-19.4$&$0.0$&$-6.0$&$-4.8$&$-81.5$&3989.8&3908.2&3934&$-$&76.7\\
		$\chi_{c0}(3860)(2{^3}P_0)$ &$0.3$&$0.0$&$-29.9$&$-2.6$&$0.0$&$-7.9$&$-40.1$&3920.5&3880.4&3852&$3862^{+26+40}_{-32-13}$&74.8\\
		$\chi_{c1}(3872)(2{^3}P_1)$ &$0.0$&$-44.0$&$-21.4$&$0.0$&$-4.7$&$-5.6$&$-73.6$&3963.3&3889.6&3925&$3871.65\pm0.06$&46.7\\
		$\chi_{c2}(3930)(2{^3}P_2)$ &$-11.5$&$-14.4$&$-25.5$&$-2.7$&$-4.7$&$-5.5$&$-64.3$&4024.5&3960.2&3972&$3922.5\pm1.0$&68.1\\ \hline
			\multicolumn{12}{c}{$\gamma=0.2$, $\beta=0.31$ GeV} \\\hline
			$h_{c}(2P)(2{^1}P_1)$ &$0.0$&$-54.4$&$-22.0$&$0.0$&$-6.9$&$-5.8$&$-89.0$&3989.8&3900.7&3934&$-$&73.8\\
			$\chi_{c0}(3860)(2{^3}P_0)$ &$-16.6$&$0.0$&$-33.2$&$-2.7$&$0.0$&$-9.4$&$-61.8$&3920.5&3858.7&3852&$3862^{+26+40}_{-32-13}$&89.0\\
			$\chi_{c1}(3872)(2{^3}P_1)$ &$0.0$&$-57.5$&$-23.6$&$0.0$&$-5.4$&$-6.7$&$-93.2$&3963.3&3870.1&3925&$3871.65\pm0.06$&43.4\\
			$\chi_{c2}(3930)(2{^3}P_2)$ &$-4.4$&$-24.6$&$-28.0$&$-2.5$&$-5.3$&$-6.5$&$-71.2$&4024.5&3953.3&3972&$3922.5\pm1.0$&75.1\\\hline
			\multicolumn{12}{c}{$\gamma=0.2$, $\beta=0.4$ GeV} \\\hline
			$h_{c}(2P)(2{^1}P_1)$ &$0.0$&$-64.5$&$-24.8$&$0.0$&$-7.9$&$-7.0$&$-104.2$&3989.8&3885.5&3934&$-$&71.1\\
			$\chi_{c0}(3860)(2{^3}P_0)$ &$-11.1$&$0.0$&$-38.8$&$-3.1$&$0.0$&$-11.5$&$-64.5$&3920.5&3856.0&3852&$3862^{+26+40}_{-32-13}$&81.0\\
			$\chi_{c1}(3872)(2{^3}P_1)$ &$0.0$&$-57.2$&$-27.3$&$0.0$&$-6.3$&$-8.1$&$-99.0$&3963.3&3864.3&3925&$3871.65\pm0.06$&60.2\\
			$\chi_{c2}(3930)(2{^3}P_2)$ &$-5.5$&$-27.7$&$-31.2$&$-2.4$&$-5.9$&$-7.8$&$-80.6$&4024.5&3943.9&3972&$3922.5\pm1.0$&84.7\\ \hline
			\multicolumn{12}{c}{$\gamma=0.2$, $\beta=0.5$ GeV} \\\hline
		$h_{c}(2P)(2{^1}P_1)$ &$0.0$&$-70.0$&$-27.8$&$0.0$&$-8.9$&$-8.1$&$-114.8$&3989.8&3875.0&3934&$-$&8.7\\
		$\chi_{c0}(3860)(2{^3}P_0)$ &$-4.6$&$0.0$&$-44.0$&$-3.5$&$0.0$&$-13.5$&$-65.6$&3920.5&3855.0&3852&$3862^{+26+40}_{-32-13}$&87.4\\
		$\chi_{c1}(3872)(2{^3}P_1)$ &$0.0$&$-58.6$&$-30.6$&$0.0$&$-7.2$&$-9.5$&$-105.9$&3963.3&3857.4&3925&$3871.65\pm0.06$&70.3\\
		$\chi_{c2}(3930)(2{^3}P_2)$ &$-8.8$&$-26.8$&$-34.5$&$-2.5$&$-6.6$&$-9.0$&$-88.1$&4024.5&3936.4&3972&$3922.5\pm1.0$&84.7\\
			\bottomrule[1pt]\bottomrule[1pt]
	\end{tabular*}}
\end{table*}

\begin{table*}[htbp]
	\caption{\label{massbeta0.25}Masses of charmonium states obtained in the adopted coupled-channel model with $\gamma=0.2$ and $\beta=0.25$ GeV. See table \ref{mass} for other explanations.}
	\scalebox{1}[1]{
		\begin{tabular*}{1.8\columnwidth}{@{\extracolsep{\fill}}c c c c c c c c c c c c c@{}}
			\toprule[1pt]\toprule[1pt]
			Meson & $D\bar{D}$ & $D\bar{D}^{*}$ & $D^*\bar{D}^*$ &$D_s\bar{D}_s$  &$D_s\bar{D}^*_s$ &$D^*_s\bar{D}^*_s$&$\delta m$&$M_A$&$M_\text{cou}$&$M_\text{non}$&$M_\text{exp}$&$P_{c\bar{c}}(\%)$  \\\hline
			$\eta_c(1S)(1{^1}S_0)$ &$0.0$&$-11.1$&$-9.8$&$0.0$&$-3.5$&$-3.1$&$-27.5$&3018.3&2990.8&2982&$2983.9\pm0.4$&97.6\\
			$J/\psi(1S)(1{^3}S_1)$ &$-2.4$&$-8.3$&$-12.7$&$-0.7$&$-2.5$&$-4.0$&$-30.7$&3136.1&3105.4&3090&$3096.900\pm0.006$&97.1\\
			$\eta_c(2S)(2{^1}S_0)$ &$0.0$&$-18.5$&$-14.4$&$0.0$&$-5.0$&$-4.2$&$-42.1$&3692.4&3650.2&3630&$3637.5\pm1.1$&92.9\\
			$\psi(2S)(2{^3}S_1)$  &$-5.4$&$-13.3$&$-17.7$&$-1.1$&$-3.5$&$-5.1$&$-46.2$&3728.9&3682.8&3672&$3686.10\pm0.06$&90.1\\
			$\eta_c(3S)(3{^1}S_0)$ &$0.0$&$-23.7$&$-26.5$&$0.0$&$-6.5$&$-5.2$&$-61.8$&4083.2&4021.4&4043&$-$&72.3\\
			$\psi(4040)(3{^3}S_1)$ &$-6.8$&$-9.9$&$-19.0$&$-1.6$&$-4.8$&$-6.4$&$-48.5$&4107.3&4058.8&4072&$4039\pm1$&82.8\\
			$\eta_c(4S)(4{^1}S_0)$ &$0.0$&$-38.3$&$-16.2$&$0.0$&$-8.3$&$-6.3$&$-69.1$&4398.8&4329.7&4384&$-$&81.6\\
			$\psi(4360)(4{^3}S_1)$ &$-5.6$&$-27.9$&$-20.0$&$-1.7$&$-5.6$&$-7.4$&$-68.2$&4417.5&4349.3&4406&$4372\pm9$&79.2\\
			$h_{c}(1P)(1{^1}P_1)$ &$0.0$&$-30.7$&$-15.7$&$0.0$&$-5.0$&$-4.2$&$-55.7$&3596.0&3540.3&3516&$3525.38\pm0.11$&91.2\\
			$\chi_{c0}(1P)(1{^3}P_0)$ &$-6.9$&$0.0$&$-23.3$&$-1.4$&$0.0$&$-6.7$&$-38.1$&3479.2&3441.1&3424&$3414.71\pm0.30$&94.6\\
			$\chi_{c1}(1P)(1{^3}P_1)$ &$0.0$&$-25.6$&$-18.2$&$0.0$&$-3.7$&$-5.1$&$-52.6$&3567.5&3514.9&3505&$3510.67\pm0.05$&92.0\\
			$\chi_{c2}(1P)(1{^3}P_2)$ &$-7.3$&$-15.5$&$-17.6$&$-1.6$&$-4.0$&$-4.5$&$-50.5$&3635.6&3585.1&3556&$3556.17\pm0.07$&91.0\\
			$h_{c}(3P)(3{^1}P_1)$ &$0.0$&$-50.2$&$-19.5$&$0.0$&$-8.0$&$-6.5$&$-84.2$&4309.4&4225.2&4279&$-$&65.4\\
			$\chi_{c0}(3P)(3{^3}P_0)$ &$-11.6$&$0.0$&$-35.7$&$-2.6$&$0.0$&$-9.6$&$-59.5$&4257.6&4198.1&4202&$-$&80.1\\
			$\chi_{c1}(3P)(3{^3}P_1)$ &$0.0$&$-34.4$&$-22.6$&$0.0$&$-6.2$&$-6.9$&$-70.2$&4283.0&4212.8&4271&$-$&71.6\\
			$\chi_{c2}(3P)(3{^3}P_2)$ &$-6.4$&$-32.5$&$-15.7$&$-2.6$&$-5.8$&$-7.1$&$-70.0$&4341.6&4271.7&4317&$-$&82.9\\
			$\eta_{c}(1D)(1{^1}D_2)$ &$0.0$&$-47.1$&$-19.7$&$0.0$&$-5.6$&$-4.5$&$-76.9$&3873.5&3796.6&3799&$-$&78.5\\
			$\psi(3770)(1{^3}D_1)$ &$-7.3$&$-6.7$&$-31.3$&$-1.2$&$-0.8$&$-7.6$&$-54.9$&3843.2&3788.3&3785&$3773.7\pm0.4$&59.6\\
			$\psi_{2}(3823)(1{^3}D_2)$ &$0.0$&$-43.9$&$-22.5$&$0.0$&$-4.6$&$-5.4$&$-76.3$&3872.2&3795.9&3800&$3823.5\pm0.5$&78.2\\
			$\psi_{3}(3842)(1{^3}D_3)$ &$-18.0$&$-22.0$&$-20.5$&$-2.3$&$-4.5$&$-4.2$&$-71.5$&3883.8&3812.3&3806&$3842.71\pm0.20$&80.9\\
			$\eta_c(2D)(2{^1}D_2)$ &$0.0$&$-43.5$&$-25.6$&$0.0$&$-7.4$&$-5.4$&$-81.9$&4203.5&4121.6&4158&$-$&85.8\\
			$\psi(4160)(2{^3}D_1)$ &$-2.7$&$3.2$&$-35.0$&$-1.2$&$-0.9$&$-9.0$&$-45.7$&4169.3&4123.6&4142&$4191\pm5$&77.9\\
			$\psi_{2}(2D)(2{^3}D_2)$ &$0.0$&$-31.3$&$-24.9$&$0.0$&$-5.9$&$-6.4$&$-68.5$&4202.2&4133.8&4158&$-$&87.9\\
			$\psi_{3}(2D)(2{^3}D_3)$ &$-9.9$&$-31.5$&$-29.9$&$-2.3$&$-6.1$&$-5.3$&$-85.0$&4214.7&4129.7&4167&$-$&73.4\\
			$\eta_{c}(3D)(3{^1}D_2)$ &$0.0$&$-67.4$&$-25.1$&$0.0$&$-7.5$&$-6.4$&$-106.4$&4491.0&4384.6&$-$&$-$&81.7\\
			$\psi(4415)(3{^3}D_1)$ &$-4.6$&$-4.5$&$-49.4$&$-1.4$&$-1.4$&$-10.4$&$-71.7$&4453.4&4381.7&$-$&$4421\pm4$&69.3\\
			$\psi_{2}(3D)(3{^3}D_2)$ &$0.0$&$-59.4$&$-33.2$&$0.0$&$-6.4$&$-7.4$&$-106.4$&4489.7&4383.2&$-$&$-$&75.1\\
			$\psi_{3}(3D)(3{^3}D_3)$ &$1.3$&$-22.8$&$-25.6$&$-4.4$&$-6.8$&$-7.4$&$-65.7$&4502.9&4437.2&$-$&$-$&67.0\\
			$h_{c}(1F)(1{^1}F_3)$ &$0.0$&$-60.6$&$-26.2$&$0.0$&$-6.0$&$-4.4$&$-97.2$&4086.7&3989.5&4026&$-$&63.3\\
			$\chi_{c2}(1F)(1{^3}F_2)$ &$2.8$&$8.4$&$-49.7$&$-0.9$&$-1.3$&$-8.4$&$-49.0$&4082.6&4033.7&4029&$-$&66.3\\
			$\chi_{c3}(1F)(1{^3}F_3)$ &$0.0$&$-42.9$&$-30.3$&$0.0$&$-5.4$&$-5.3$&$-83.8$&4088.1&4004.2&4029&$-$&56.4\\
			$\chi_{c4}(1F)(1{^3}F_4)$ &$-7.8$&$-37.2$&$-29.7$&$-3.2$&$-4.9$&$-3.9$&$-86.7$&4087.3&4000.6&4021&$-$&67.5\\
			\hline
			\bottomrule[1pt]\bottomrule[1pt]
	\end{tabular*}}
\end{table*}

In the adopted model, $\gamma$ and $\beta$ are two important parameters. Now we briefly discuss the effects of varying $\gamma$ and $\beta$ on the mass spectra of charmonium states. If one adopts a larger (smaller) $\gamma$, the coupled-channel effects would become more (less) important and the charmonium masses become smaller (larger). In the above discussions, we fixed $\gamma=0.2$ which are extracted from the measured widths of $\psi(3770)$ and $\chi_{c2}(3930)$.
Note that larger $\gamma$'s may be adopted to understand the meson spectra in the literature \cite{Deng:2023mza}, but this does not mean that our value of $\gamma$ is too small. A suppressed factor is probably introduced when one uses a larger $\gamma$. The tendency of their results is similar to that of our calculations.

To see the variation effect on the charmonium state when we take a different $\beta$, one considers the $\chi_{c1}(3872)$ as an example. We change the value of $\beta$ in the range $0.25\sim0.50$ GeV from the adopted $\beta=0.31$ GeV in the above calculations. The numerical results for the four $2P$ $c\bar{c}$ states are summarized in table \ref{diffbeta}. From the table, the mass of $\chi_{c1}(2P)$ is slightly smaller than the experimental value and the coupled-channel effects are somewhat increased for a larger $\beta$. Meanwhile, the mass shift contribution from the $D\bar{D}^*$ channel remains almost unchanged. This indicates that the mass of $\chi_{c1}(2P)$ is relatively stable with the increase of $\beta$, although the proportion of the $c\bar{c}$ component changes significantly. However, when a 0.06 GeV smaller value $\beta=0.25$ GeV is used, the resulting mass of $\chi_{c1}(2P)$ is about 20 MeV larger than the case of $\beta=0.31$ GeV. The mass shift contributions from all the relevant channels are reduced and the $c\bar{c}$ proportion is just slightly larger. From table \ref{diffbeta}, it is clear that variations for the masses of $h_c(2P)$, $\chi_{c0}(2P)$, and $\chi_{c2}(2P)$ are all compatible with the above feature that a smaller (larger) $\beta$ leads to a larger (smaller) charmonium mass, but the contributions to the mass shift from different channels do not show a unique trend. The possibility of the $c\bar{c}$ component may change significantly with some special $\beta$, e.g. the case of $h_c(2P)$ around $\beta=0.5$ GeV. In the general case, it is difficult to find a unique feature. It is also possible that a smaller (larger) $\beta$ leads to a smaller (larger) charmonium mass and a smaller (larger) $P_{c\bar{c}}$. To see the dependence of the charmonium properties on the value of $\beta$, we present numerical results for other states with $\beta=0.25$ GeV, 0.4 GeV, and 0.5 GeV in tables \ref{massbeta0.25}-\ref{massbeta0.5}.

In Ref. \cite{Duan:2020tsx}, the authors noted the $\beta$-induced node effects on the mass and decay width of the radially excited $\chi_{c0}(2P)$. Here, based on our numerical results with different $\beta$'s, one sees that the existence of nodes and coupled channel effects for various radially excited $c\bar{c}$ states leads to the complicated properties of charmonium states and the difficulty to identify charmonia from other structures or effects.

Although we improve the nonrelativistic model by considering coupled-channel effects, uncertainties in getting the mass spectrum remain. As a result, it is difficult to give masses compatible with all the measured values.  For example, the obtained mass of $\psi(2D)$ is about 60 MeV smaller than the experimental result. The relativistic effect is probably a source to understand the differences. This issue and other possible reasons will be further explored in future works.

\begin{table*}[!htbp]
\caption{\label{massbeta0.4}Masses of charmonium states obtained in the adopted coupled-channel model with $\gamma=0.2$ and $\beta=0.4$ GeV. See table \ref{mass} for other explanations.}
\scalebox{1}[1]{
\begin{tabular*}{1.8\columnwidth}{@{\extracolsep{\fill}}c c c c c c c c c c c c c@{}}
	\toprule[1pt]\toprule[1pt]
	Meson & $D\bar{D}$ & $D\bar{D}^{*}$ & $D^*\bar{D}^*$ &$D_s\bar{D}_s$  &$D_s\bar{D}^*_s$ &$D^*_s\bar{D}^*_s$&$\delta m$&$M_A$&$M_\text{cou}$&$M_\text{non}$&$M_\text{exp}$&$P_{c\bar{c}}(\%)$  \\\hline
	
$\eta_c(1S)(1{^1}S_0)$ &$0.0$&$-23.1$&$-20.9$&$0.0$&$-7.3$&$-6.7$&$-58.0$&3018.3&2960.4&2982&$2983.9\pm0.4$&95.9\\
$J/\psi(1S)(1{^3}S_1)$ &$-4.8$&$-16.9$&$-26.6$&$-1.4$&$-5.2$&$-8.4$&$-63.3$&3136.1&3072.8&3090&$3096.900\pm0.006$&95.2\\
$\eta_c(2S)(2{^1}S_0)$ &$0.0$&$-27.7$&$-23.4$&$0.0$&$-7.9$&$-7.1$&$-66.2$&3692.4&3626.2&3630&$3637.5\pm1.1$&92.3\\
$\psi(2S)(2{^3}S_1)$  &$-6.6$&$-19.5$&$-28.4$&$-1.6$&$-5.4$&$-8.6$&$-70.0$&3728.9&3658.9&3672&$3686.10\pm0.06$&90.6\\
$\eta_c(3S)(3{^1}S_0)$ &$0.0$&$-19.5$&$-27.1$&$0.0$&$-7.7$&$-6.8$&$-61.1$&4083.2&4022.1&4043&$-$&76.7\\
$\psi(4040)(3{^3}S_1)$ &$-5.1$&$-14.5$&$-32.2$&$-1.2$&$-5.3$&$-8.1$&$-66.4$&4107.3&4040.9&4072&$4039\pm1$&82.4\\
$\eta_c(4S)(4{^1}S_0)$ &$0.0$&$-22.5$&$-17.4$&$0.0$&$-6.9$&$-6.4$&$-53.2$&4398.8&4345.6&4384&$-$&92.8\\
$\psi(4360)(4{^3}S_1)$ &$-3.8$&$-15.5$&$-21.1$&$-1.2$&$-4.6$&$-7.6$&$-53.7$&4417.5&4363.8&4406&$4372\pm9$&92.4\\
$h_{c}(1P)(1{^1}P_1)$ &$0.0$&$-51.1$&$-28.2$&$0.0$&$-8.8$&$-7.9$&$-95.9$&3596.0&3500.1&3516&$3525.38\pm0.11$&89.4\\
$\chi_{c0}(1P)(1{^3}P_0)$ &$-12.8$&$0.0$&$-42.8$&$-2.5$&$0.0$&$-12.7$&$-70.8$&3479.2&3408.4&3424&$3414.71\pm0.30$&92.4\\
$\chi_{c1}(1P)(1{^3}P_1)$ &$0.0$&$-44.5$&$-32.1$&$0.0$&$-6.7$&$-9.5$&$-92.7$&3567.5&3474.8&3505&$3510.67\pm0.05$&90.0\\
$\chi_{c2}(1P)(1{^3}P_2)$ &$-9.9$&$-24.4$&$-32.1$&$-2.6$&$-6.8$&$-8.3$&$-84.2$&3635.6&3551.4&3556&$3556.17\pm0.07$&90.1\\
$h_{c}(3P)(3{^1}P_1)$ &$0.0$&$-45.9$&$-18.4$&$0.0$&$-6.9$&$-6.9$&$-78.2$&4309.4&4231.2&4279&$-$&87.7\\
$\chi_{c0}(3P)(3{^3}P_0)$ &$-10.6$&$0.0$&$-29.2$&$-2.3$&$0.0$&$-10.9$&$-53.0$&4257.6&4204.6&4202&$-$&93.1\\
$\chi_{c1}(3P)(3{^3}P_1)$ &$0.0$&$-40.2$&$-20.6$&$0.0$&$-6.0$&$-7.6$&$-74.4$&4283.0&4208.5&4271&$-$&85.2\\
$\chi_{c2}(3P)(3{^3}P_2)$ &$-4.6$&$-17.1$&$-22.6$&$-1.9$&$-4.8$&$-7.3$&$-58.5$&4341.6&4283.2&4317&$-$&89.4\\
$\eta_{c}(1D)(1{^1}D_2)$ &$0.0$&$-58.7$&$-29.6$&$0.0$&$-8.4$&$-7.3$&$-104.0$&3873.5&3769.5&3799&$-$&83.2\\
$\psi(3770)(1{^3}D_1)$ &$-23.0$&$-9.1$&$-44.9$&$-1.7$&$-1.3$&$-12.2$&$-92.2$&3843.2&3751.0&3785&$3773.7\pm0.4$&78.4\\
$\psi_{2}(3823)(1{^3}D_2)$ &$0.0$&$-55.6$&$-32.8$&$0.0$&$-6.9$&$-8.7$&$-104.0$&3872.2&3768.2&3800&$3823.5\pm0.5$&82.7\\
$\psi_{3}(3842)(1{^3}D_3)$ &$-13.1$&$-26.0$&$-32.2$&$-2.9$&$-6.7$&$-7.0$&$-87.9$&3883.8&3795.8&3806&$3842.71\pm0.20$&85.6\\
$\eta_c(2D)(2{^1}D_2)$ &$0.0$&$-47.8$&$-23.9$&$0.0$&$-8.0$&$-6.6$&$-86.3$&4203.5&4117.2&4158&$-$&97.9\\
$\psi(4160)(2{^3}D_1)$ &$-7.0$&$-6.6$&$-41.1$&$-1.0$&$-1.7$&$-10.7$&$-68.3$&4169.3&4101.0&4142&$4191\pm5$&86.4\\
$\psi_{2}(2D)(2{^3}D_2)$ &$0.0$&$-45.8$&$-28.9$&$0.0$&$-7.0$&$-7.6$&$-89.4$&4202.2&4112.8&4158&$-$&91.7\\
$\psi_{3}(2D)(2{^3}D_3)$ &$-5.9$&$-16.2$&$-22.1$&$-2.5$&$-6.1$&$-6.8$&$-59.5$&4214.7&4155.2&4167&$-$&87.3\\
$\eta_{c}(3D)(3{^1}D_2)$ &$0.0$&$-38.7$&$-20.4$&$0.0$&$-6.4$&$-6.1$&$-71.7$&4491.0&4419.3&$-$&$-$&93.8\\
$\psi(4415)(3{^3}D_1)$ &$-4.7$&$-5.4$&$-31.9$&$-1.3$&$-1.2$&$-9.7$&$-54.2$&4453.4&4399.2&$-$&$4421\pm4$&92.6\\
$\psi_{2}(3D)(3{^3}D_2)$ &$0.0$&$-36.9$&$-23.2$&$0.0$&$-5.5$&$-6.9$&$-72.4$&4489.7&4417.3&$-$&$-$&94.9\\
$\psi_{3}(3D)(3{^3}D_3)$ &$-5.5$&$-14.8$&$-21.9$&$-2.1$&$-5.0$&$-6.2$&$-55.4$&4502.9&4447.5&$-$&$-$&93.2\\
$h_{c}(1F)(1{^1}F_3)$ &$0.0$&$-68.1$&$-28.9$&$0.0$&$-7.4$&$-6.3$&$-110.7$&4086.7&3976.0&4026&$-$&80.3\\
$\chi_{c2}(1F)(1{^3}F_2)$ &$-1.7$&$-15.1$&$-45.7$&$-2.0$&$-1.4$&$-11.1$&$-77.0$&4082.6&4005.6&4029&$-$&80.8\\
$\chi_{c3}(1F)(1{^3}F_3)$ &$0.0$&$-67.4$&$-30.9$&$0.0$&$-6.4$&$-7.2$&$-111.9$&4088.1&3976.1&4029&$-$&81.6\\
$\chi_{c4}(1F)(1{^3}F_4)$ &$-14.1$&$-26.3$&$-33.0$&$-2.8$&$-5.9$&$-5.7$&$-87.9$&4087.3&3999.4&4021&$-$&80.9\\
\hline
			\bottomrule[1pt]\bottomrule[1pt]
	\end{tabular*}}
\end{table*}

\begin{table*}[!htbp]
\caption{\label{massbeta0.5}Masses of charmonium states obtained in the adopted coupled-channel model with $\gamma=0.2$ and $\beta=0.5$ GeV. See table \ref{mass} for other explanations.}
\scalebox{1}[1]{
\begin{tabular*}{1.8\columnwidth}{@{\extracolsep{\fill}}c c c c c c c c c c c c c@{}}
	\toprule[1pt]\toprule[1pt]
	Meson & $D\bar{D}$ & $D\bar{D}^{*}$ & $D^*\bar{D}^*$ &$D_s\bar{D}_s$  &$D_s\bar{D}^*_s$ &$D^*_s\bar{D}^*_s$&$\delta m$&$M_A$&$M_\text{cou}$&$M_\text{non}$&$M_\text{exp}$&$P_{c\bar{c}}(\%)$  \\\hline
	
$\eta_c(1S)(1{^1}S_0)$ &$0.0$&$-31.2$&$-28.7$&$0.0$&$-9.9$&$-9.2$&$-79.1$&3018.3&2939.2&2982&$2983.9\pm0.4$&95.1\\
$J/\psi(1S)(1{^3}S_1)$ &$-6.3$&$-22.6$&$-36.1$&$-1.9$&$-7.1$&$-11.5$&$-85.4$&3136.1&3050.7&3090&$3096.900\pm0.006$&94.3\\
$\eta_c(2S)(2{^1}S_0)$ &$0.0$&$-32.8$&$-28.5$&$0.0$&$-9.6$&$-8.8$&$-79.7$&3692.4&3612.7&3630&$3637.5\pm1.1$&92.3\\
$\psi(2S)(2{^3}S_1)$  &$-7.2$&$-22.8$&$-34.4$&$-1.8$&$-6.5$&$-10.5$&$-83.3$&3728.9&3645.6&3672&$3686.10\pm0.06$&91.0\\
$\eta_c(3S)(3{^1}S_0)$ &$0.0$&$-19.1$&$-27.5$&$0.0$&$-8.4$&$-7.6$&$-62.6$&4083.2&4020.6&4043&$-$&82.3\\
$\psi(4040)(3{^3}S_1)$ &$-4.4$&$-12.4$&$-34.0$&$-1.4$&$-5.7$&$-9.0$&$-67.0$&4107.3&4040.3&4072&$4039\pm1$&95.3\\
$\eta_c(4S)(4{^1}S_0)$ &$0.0$&$-17.8$&$-18.0$&$0.0$&$-7.1$&$-6.6$&$-49.6$&4398.8&4349.2&4384&$-$&98.2\\
$\psi(4360)(4{^3}S_1)$ &$-3.5$&$-12.0$&$-21.0$&$-1.2$&$-4.8$&$-7.8$&$-50.3$&4417.5&4367.2&4406&$4372\pm9$&97.6\\
$h_{c}(1P)(1{^1}P_1)$ &$0.0$&$-62.7$&$-35.7$&$0.0$&$-11.1$&$-10.1$&$-119.6$&3596.0&3476.4&3516&$3525.38\pm0.11$&89.0\\
$\chi_{c0}(1P)(1{^3}P_0)$ &$-16.8$&$0.0$&$-54.6$&$-3.3$&$0.0$&$-16.4$&$-91.0$&3479.2&3388.2&3424&$3414.71\pm0.30$&91.7\\
$\chi_{c1}(1P)(1{^3}P_1)$ &$0.0$&$-55.9$&$-40.1$&$0.0$&$-8.5$&$-12.1$&$-116.6$&3567.5&3450.9&3505&$3510.67\pm0.05$&89.4\\
$\chi_{c2}(1P)(1{^3}P_2)$ &$-11.2$&$-29.1$&$-41.1$&$-3.1$&$-8.4$&$-10.7$&$-103.7$&3635.6&3531.9&3556&$3556.17\pm0.07$&90.1\\
$h_{c}(3P)(3{^1}P_1)$ &$0.0$&$-35.2$&$-18.9$&$0.0$&$-7.5$&$-7.3$&$-68.9$&4309.4&4240.4&4279&$-$&91.6\\
$\chi_{c0}(3P)(3{^3}P_0)$ &$-7.6$&$0.0$&$-31.3$&$-2.6$&$0.0$&$-11.7$&$-53.2$&4257.6&4204.4&4202&$-$&94.4\\
$\chi_{c1}(3P)(3{^3}P_1)$ &$0.0$&$-30.2$&$-21.4$&$0.0$&$-6.2$&$-8.2$&$-66.1$&4283.0&4216.9&4271&$-$&93.9\\
$\chi_{c2}(3P)(3{^3}P_2)$ &$-6.0$&$-15.6$&$-20.3$&$-1.7$&$-5.1$&$-7.6$&$-56.4$&4341.6&4285.2&4317&$-$&95.7\\
$\eta_{c}(1D)(1{^1}D_2)$ &$0.0$&$-64.7$&$-34.7$&$0.0$&$-9.9$&$-8.9$&$-118.1$&3873.5&3755.4&3799&$-$&85.3\\
$\psi(3770)(1{^3}D_1)$ &$-20.4$&$-10.7$&$-52.4$&$-2.0$&$-1.6$&$-14.8$&$-101.9$&3843.2&3741.4&3785&$3773.7\pm0.4$&77.3\\
$\psi_{2}(3823)(1{^3}D_2)$ &$0.0$&$-61.7$&$-38.1$&$0.0$&$-8.2$&$-10.4$&$-118.4$&3872.2&3753.8&3800&$3823.5\pm0.5$&84.8\\
$\psi_{3}(3842)(1{^3}D_3)$ &$-12.5$&$-27.9$&$-38.4$&$-3.2$&$-7.7$&$-8.6$&$-98.4$&3883.8&3785.4&3806&$3842.71\pm0.20$&88.1\\
$\eta_c(2D)(2{^1}D_2)$ &$0.0$&$-40.5$&$-25.4$&$0.0$&$-8.4$&$-7.3$&$-81.6$&4203.5&4121.9&4158&$-$&93.5\\
$\psi(4160)(2{^3}D_1)$ &$-7.1$&$-3.3$&$-41.1$&$-1.1$&$-1.8$&$-11.8$&$-66.2$&4169.3&4103.1&4142&$4191\pm5$&90.4\\
$\psi_{2}(2D)(2{^3}D_2)$ &$0.0$&$-35.5$&$-29.4$&$0.0$&$-7.4$&$-8.4$&$-80.7$&4202.2&4121.5&4158&$-$&92.2\\
$\psi_{3}(2D)(2{^3}D_3)$ &$-6.6$&$-20.7$&$-22.8$&$-2.6$&$-6.1$&$-7.3$&$-66.2$&4214.7&4148.5&4167&$-$&92.8\\
$\eta_{c}(3D)(3{^1}D_2)$ &$0.0$&$-33.1$&$-17.3$&$0.0$&$-6.5$&$-6.3$&$-63.2$&4491.0&4427.8&$-$&$-$&94.8\\
$\psi(4415)(3{^3}D_1)$ &$-4.3$&$-3.3$&$-27.6$&$-1.3$&$-1.3$&$-10.2$&$-47.9$&4453.4&4405.5&$-$&$4421\pm4$&95.8\\
$\psi_{2}(3D)(3{^3}D_2)$ &$0.0$&$-30.0$&$-19.8$&$0.0$&$-5.6$&$-7.2$&$-62.7$&4489.7&4427.0&$-$&$-$&94.5\\
$\psi_{3}(3D)(3{^3}D_3)$ &$-6.0$&$-15.0$&$-17.2$&$-1.9$&$-4.8$&$-6.3$&$-51.1$&4502.9&4451.8&$-$&$-$&96.5\\
$h_{c}(1F)(1{^1}F_3)$ &$0.0$&$-63.7$&$-30.9$&$0.0$&$-8.1$&$-7.3$&$-110.0$&4086.7&3976.7&4026&$-$&83.1\\
$\chi_{c2}(1F)(1{^3}F_2)$ &$-9.8$&$-15.6$&$-47.1$&$-1.7$&$-1.6$&$-12.5$&$-88.3$&4082.6&3994.3&4029&$-$&87.4\\
$\chi_{c3}(1F)(1{^3}F_3)$ &$0.0$&$-63.0$&$-33.0$&$0.0$&$-7.0$&$-8.3$&$-111.3$&4088.1&3976.8&4029&$-$&83.0\\
$\chi_{c4}(1F)(1{^3}F_4)$ &$-12.0$&$-24.7$&$-34.7$&$-2.9$&$-6.4$&$-6.6$&$-87.3$&4087.3&4000.0&4021&$-$&86.2\\
\hline
			\bottomrule[1pt]\bottomrule[1pt]
	\end{tabular*}}
\end{table*}

\section*{Acknowledgments}

Z.L. Man thanks Dr. S.Q. Luo for helpful discussions. This project was supported by the National Natural Science Foundation of China under Grant Nos. 12235008,  11875226, and by the Shandong Province Natural Science Foundation (ZR202210240019).

\end{document}